\documentclass[aps,prd,longbibliography,nofootinbib]{revtex4-2}

\usepackage{graphicx}
\usepackage{amsmath,amssymb,bm}
\usepackage{booktabs}
\usepackage{tikz}
\usetikzlibrary{arrows.meta,positioning}
\usepackage{placeins}
\graphicspath{{figures/}}

\newcommand{\Cvar}{C_{\rm var}}
\newcommand{\Eavg}{\mathbb E}
\usepackage{microtype}
\usepackage{hyperref}
\hypersetup{hidelinks,pdftitle={An Analytic Error-Budget Model for the Habitable Worlds Observatory Coronagraph: From Optical Disturbances to Flux-Ratio Noise},pdfauthor={Slava G. Turyshev}}
\usepackage{hyperref}
\hypersetup{colorlinks=true,allcolors=blue}

\newcommand{\FRN}{\mathrm{FRN}}
\newcommand{\WFE}{\mathrm{WFE}}
\newcommand{\E}{\mathcal{E}}          

\newcommand{\Ecoh}{\E_{\rm coh}}

\newcommand{\I}{\mathcal{I}}          
\newcommand{\Craw}{C_{\rm raw}}
\newcommand{\Ccoh}{C_{\rm coh}}
\newcommand{\Cinc}{C_{\rm inc}}
\newcommand{\Cstab}{C_{\rm stab}}
\newcommand{\kappac}{\kappa_c}

\newcommand{\lam}{\lambda}

\newcommand{\Dtel}{D}
\newcommand{\OWA}{\mathrm{OWA}}
\newcommand{\IWA}{\mathrm{IWA}}
\newcommand{\Ree}{\mathrm{Re}}

\newcommand{\avg}[1]{\left\langle #1 \right\rangle}
\newcommand{\Var}{\mathrm{Var}}
\newcommand{\Cov}{\mathrm{Cov}}

\newcommand{\Tr}{\mathrm{Tr}}
\newcommand{\diag}{\mathrm{diag}}
\newcommand{\sinc}{\mathrm{sinc}}
\newcommand{\Cop}{\mathcal{C}}

\newcommand{\Mpp}{M_{\rm pp}}
\newcommand{\etapp}{\eta_{\rm pp}}

\newcommand{\Cincstat}{C_{\rm inc,stat}}
\newcommand{\Cphot}{C_{\rm phot}}

\newcommand{\SigmaDx}{\bm\Sigma_{\Delta x}}
\newcommand{\SigmaDI}{\bm\Sigma_{\Delta I}}
\newcommand{\SigmaInc}{\bm\Sigma_{\Delta I,{\rm inc}}}

\newcommand{\Hmat}{\bm H}

\newcommand{\Amat}{\bm A}

\newcommand{\Mmode}{m}
\newcommand{\Dmode}{d}

\begin{document}

\title{An Analytic Error-Budget Model for the Habitable Worlds Observatory Coronagraph: \\ From Optical Disturbances to Flux-Ratio Noise}

\author{Slava G. Turyshev}
\affiliation{
Jet Propulsion Laboratory, California Institute of Technology,\\
4800 Oak Grove Drive, Pasadena, CA 91109-0899, USA
}%
\date{\today}

\begin{abstract}
A coronagraph stability allocation must protect a planet measurement, not merely satisfy a contrast target. Residual starlight both generates photon noise and interferes with small field perturbations, coupling optical stability to detection reliability. We develop a modular analytic model for the Habitable Worlds Observatory (HWO), using a reduced two-roll angular-differential-imaging realization of its OS-1 scenario. Complex-field sensitivities, wavefront sensing and control, finite observing windows, polarization, and detector calibration are propagated to flux-ratio noise (FRN) and detection probabilities. Joint-visit Gaussian quadratic moments retain differential bias and cross covariance; spatial overlap and field phase determine which disturbances resemble a planet. A declared 6-m, 500-nm benchmark shows why the science test must precede category allocation. For a 115.46-parts-per-trillion (ppt) quadrature planet, 30,000 independent Gaussian search statistics, and a $10^{-3}$ family-wise false-alarm objective, 20-ppt FRN gives 64.55\% conditional detection power; 99\% requires 14.94 ppt. Photon and calibration noise exhaust the optical remainder at 8.11 pc in 100 h under the fixed response, so tighter stability alone cannot restore feasibility. A complementary intermittency test preserves disturbance mean and covariance while changing detection tails. An unrecognized one-percent high-variance component carrying twenty percent of the variance tightens the modal allowance by a factor of 2.88; an ideal, exactly known state label reduces that penalty to 1.29 under equal conditional false-alarm limits and the same average-power objective. Thus requirements depend on both the optical response and the information retained by the observing protocol. The framework identifies when suppression, stability, calibration, or state monitoring addresses the limiting mechanism. Numerical tolerances are controlled model results; flight allocation requires architecture-specific responses and calibrated disturbance and state-estimation laws.
\end{abstract}

\maketitle

{\small\tableofcontents}


\section{Introduction}
\label{sec:scope}

Reflected-light imaging of habitable-zone terrestrial planets provides the observational basis for investigating their atmospheres and physical properties. For the Habitable Worlds Observatory (HWO), the ability to acquire reflected-light spectra therefore begins with a more basic requirement: a planet must be distinguished reliably from residual stellar light at its location \cite{Astro2020,NASA2026HWO,Liu2026EAC}. A deep coronagraphic dark hole is essential, but its mean intensity does not determine whether that distinction can be made over a finite observing program. The measurement also depends on how the residual image changes, how those changes are calibrated, and how much planet signal the extraction retains.

The physical distinction is between photons already admitted to the measurement and changes in the optical field that admits them. Mean stellar leakage contributes shot noise even when its expected image is subtracted perfectly. The residual coherent field can also interfere with starlight redistributed by an optical perturbation, producing an intensity change linear in disturbance amplitude. A repeatable change between observations produces a flux bias, whereas fluctuations about that change produce covariance. Their spatial patterns matter because the same disturbance can cancel in one differential observable and resemble a planet in another. Consequently, a wavefront-error (WFE) root-mean-square (RMS) target becomes a science requirement only after specifying the optical response, observing sequence, and calibrated estimator.

The central question is how to translate a planet-detection objective into disturbance allocations without losing these physical distinctions. Flux-ratio noise (FRN), expressed in parts per trillion (ppt), provides a common uncertainty scale, but its acceptable value follows from the planet brightness and the detection test. Searching many locations requires rejection of rare stellar residuals, not merely a large mean signal-to-noise ratio. We therefore examine both the photon and stability terms that set the width of the flux distribution and the uncommon observing states that can determine its detection tails.

Within the existing HWO engineering framework, we retain wavefront, pointing, pupil-shear, beamwalk, polarization, detector, and calibration terms, together with wavefront sensing and control (WFSC) and power-spectral-density (PSD) propagation. The purpose is to determine which reduced descriptions remain adequate for the measured planet signal. Reducing mean leakage, improving inter-visit repeatability, and characterizing rare residuals address different limitations; the error budget must distinguish their scientific leverage.

\subsection{HWO mission context and representative architecture used for budgeting}
\label{sec:hwo_context}

HWO's Exploratory Analytic Cases (EACs) provide a systems-engineering setting in which observatory stability, coronagraph response, and observing operations are developed together \cite{Liu2026EAC}. The representative context here is an off-axis, segmented, approximately 6-m-class telescope with two deformable mirrors (DMs), low-order wavefront sensing (LOWFS), and high-order wavefront sensing and control (HOWFS/HOWFSC) \cite{Feinberg2026Maturation,Tesch2026WFSC}. These elements determine the disturbance families that the budget must accommodate. They do not, by themselves, supply the numerical sensitivities used in an allocation; those are specified separately for each optical and control state.


Tables~\ref{tab:nomenclature_core} and \ref{tab:nomenclature_model} define the notation; $1~\mathrm{ppt}=10^{-12}$ in flux ratio. Charge-6 vortex studies illustrate the architecture dependence of tolerancing \cite{Ruane2017VortexSegmented}. The scalar sensitivity used below is a declared test input, while its identification with an EAC mode requires coordinate-normalized complex maps. Similarly, the approximately 20-ppt scale discussed in HWO flowdown studies \cite{NematiStahl2024HWOErrorBudgetRoman} is a useful reference, but the requirement for a particular planet follows from its flux and the specified detection test.

\paragraph*{Driving-case formulation for allocations.}
The broadband and narrow-band allocations in this paper should be understood as tied to a specified science-driving target/mode pair, rather than to an unspecified target list. Let
\begin{equation}
\mathcal D_{\rm drv}\equiv
\left\{
\lambda_c,\,
\Delta\lambda,\,
\rho_{\rm obs},\,
f_{p,\rm req},\,
\mathrm{SNR}_{\rm req},\,
t_{\rm alloc}
\right\}_{\rm drv}
\label{eq:driving_case}
\end{equation}
denote the corresponding driving case. The top-level requirement is then
\begin{equation}
\FRN_{\rm req}^{\rm (drv)}(\lambda_c)=\frac{f_{p,\rm req}(\lambda_c)}{\mathrm{SNR}_{\rm req}},
\label{eq:frn_req_drv}
\end{equation}
with raw leakage, stability, throughput, and covariance evaluated for the same $\mathcal D_{\rm drv}$. Visible and near-infrared (NIR) bands and imaging and spectroscopy objectives therefore require their own allocations, rather than a common architecture-invariant contrast number.

\subsection{Architecture- and observing-mode indices}
\label{sec:mode_indices}

A change of coronagraph or observing strategy changes the measurement, not merely a label on its noise budget. To keep such comparisons consistent, optical responses, control operating points, and observing differentials carry explicit mode labels:
\begin{equation}
\left\{
\E_0,\,
J_k,\,
\Ccoh,\,
\Cinc,\,
\Cstab,\,
\tau_{\rm core},\,
\kappac,\,
\bm\Sigma
\right\}
\rightarrow
\left\{
\E_0^{(\Mmode,\Dmode)},\,
J_k^{(\Mmode,\Dmode)},\,
\Ccoh^{(\Mmode,\Dmode)},\,
\Cinc^{(\Mmode,\Dmode)},\,
\Cstab^{(\Mmode,\Dmode)},\,
\tau_{\rm core}^{(\Mmode,\Dmode)},\,
\kappac^{(\Mmode,\Dmode)},\,
\bm\Sigma^{(\Mmode,\Dmode)}
\right\},
\label{eq:mode_indexing}
\end{equation}
where $\Mmode$ labels the coronagraph / DM operating point and $\Dmode$ labels the observing
differential (RDI, roll pair, ADI, revisit, \emph{etc.}).
Whenever annulus averages are used, the region $R$ or working-angle interval must also be explicit.

\paragraph*{Scope of the present numerical examples.}
OS-1 is the HWO project scenario described by Liu et al., derived from Roman OS-11 \cite{Liu2026EAC}. Here it is reduced to two angular-differential-imaging (ADI) roll states with both nonoverlapping planet lobes retained. The full project sequence also visits a brighter star to establish and maintain the dark hole; our declared live fraction includes overhead but does not simulate that complete sequence. The project excludes reference differential imaging (RDI) subtraction because stellar color and angular-size mismatch bias the comparison. RDI is retained only as a resource-accounting comparator, not as a claimed HWO baseline.

Coronagraph observing and post-processing (COPP) choices enter the mode index through
\begin{equation}
m=(m_{\rm mask},m_{\rm band},m_{\rm WFSC},m_{\rm COPP}),\label{eq:mode_indexing_expanded}
\end{equation}
with ADI as the numerical baseline. Vortex, phase-induced amplitude apodization (PIAA), and photonic architectures require their own fields, throughput, detector response, and observing cost before comparison.

\subsection{Roman CGI heritage and key differences relative to HWO}
\label{sec:roman_heritage}

Roman's Coronagraph Instrument (CGI) provides the analytical and end-to-end modeling heritage linking FRN, coherent-field interference, photometric noise, stability, and operations \cite{Nemati2023RomanErrorBudget,Krist2023ModelingRoman,TangaPoberezhskiy2021RomanCGIOps}. HWO flowdown and mirror-scatter studies extend that foundation \cite{NematiStahl2024HWOErrorBudgetRoman,NematiStahl2025ScatterFlowdown}. The closest systems-level antecedent is Liu et al.\ \cite{Liu2026EAC}: their Roman-derived observing scenario, complex mean field and temporal variance together with their inter-visit changes, and disturbance taxonomy are the starting point of this paper.

Other approaches address complementary parts of the same problem. The Pair-based Analytical model for Segmented Telescopes Imaging from Space (PASTIS) supplies matrix-based tolerancing, with experimental validation and extensions to segment thermal and sensing behavior \cite{Laginja2021PASTIS,Laginja2022Validation,Sahoo2026Thermal}. Error Budget Software (EBS) connects disturbance and WFSC sensitivities to exposure calculations, while Flagey et al. formulate a PASTIS-based multiscale HWO budget \cite{Steiger2026EBS,Flagey2026Budget}. The need for calibrated detection statistics is also established: small-sample effects alter high-contrast imaging thresholds \cite{Mawet2014SmallSampleStats}, and Dannert et al. derive non-Gaussian differential-output statistics and detection tests for nulling interferometers \cite{Dannert2025NonGaussian}.

The issue addressed here is what must be retained when these physical and statistical descriptions are reduced to category allocations and passed between models. Neither a scalar contrast sensitivity nor an estimator variance contains all the information required for that transfer. We combine joint-visit quadratic moments [Eq.~\eqref{eq:gaussian_moments}], distinct spatial-overlap and phase factors [Eq.~\eqref{eq:phase_response}], finite-window averaging, and a common count normalization. Controlled comparisons keep the target, response, and observing cost fixed while changing a single reduction or distributional assumption. This makes their consequences visible as changes in allowable disturbances, exposure time, or detection probability.

The contribution is thus a tested allocation-to-estimator connection, rather than a new Gaussian identity, a new definition of FRN, or a claim that existing tools omit covariance. Its most consequential test preserves disturbance mean and covariance while solving the false-alarm and detection-power constraints with exact conditional-Poisson statistics. Together with the target-dependent boundaries and model products in Table~\ref{tab:products_schema}, this provides a way to decide whether a proposed simplification is adequate for the intended science measurement.

\subsection{Interface between integrated modeling, allocations, and science estimators}
\label{sec:interfaces}

The analysis follows the measurement in the forward direction: optical and control models determine the stellar counts and their statistics; calibration and extraction turn those counts into a flux estimate. Requirement flowdown then runs in the reverse direction. A specified planet and detection objective set the allowed estimator error, which is apportioned among physical disturbances using the same responses and observing cost. Keeping these two directions consistent is the organizing principle of the paper.

Section~\ref{sec:frn} defines FRN, planet geometry, and photometric normalization. Sections~\ref{sec:raw_contrast}--\ref{sec:stability} develop the optical moments and disturbance allocations; Secs.~\ref{sec:temporal} and \ref{sec:pol} add temporal and polarization response. Section~\ref{sec:closure} applies the framework to a common count budget and target-dependent boundaries, using the estimators and detection test developed in Sec.~\ref{sec:signalextraction}. Section~\ref{sec:validation} tests optical and distributional assumptions, compares unrecognized and exactly labeled observing states, and interprets the design implications. Section~\ref{sec:concl} synthesizes the results. Appendices~\ref{app:moments}--\ref{app:tails} give the moment derivation, count verification, numerical procedures, and distribution inversion.

The closed-form moments are exact for the stated Gaussian polynomial model, while their optical application requires a validated field expansion. The numerical examples use a blackbody star, Lambertian planet, fixed instrumental response, and stationary-equivalent observing time. The 5-pc case is a favorable benchmark, not the project's driving target; overhead is charged through a live fraction rather than a simulation of the full OS-1 sequence. These choices isolate the allocation logic. Architecture-specific maps, finite-star and vector responses, control histories, and calibrated residual distributions are required to obtain flight tolerances.

\begin{table*}[t]
\caption{Core optical, geometric, and science-facing notation used throughout the paper.}
\label{tab:nomenclature_core}
\begin{tabular}{p{0.12\linewidth} p{0.58\linewidth} p{0.18\linewidth}}
\toprule
Symbol & Meaning & Typical units \\
\midrule
$\lam$ & Wavelength & nm or $\mu$m \\
$\Dtel$ & Telescope diameter & m \\
$\rho$ & Working angle in units of $\lam/\Dtel$ & dimensionless \\
$\IWA,\OWA$ & Inner and outer working angles & $\lam/\Dtel$ \\
$f_p(\lam)$ & Planet-to-star flux ratio & dimensionless \\
$\Phi(\alpha)$ & Planet phase function & dimensionless \\
$\tau_{\rm core}(\lam,\rho)$ & Planet core throughput & dimensionless \\
$\Omega_{\rm core}$ & Solid angle of the adopted core aperture & sr \\
$C_\star(\lam)$ & Stellar count rate before coronagraph suppression & e$^-\,{\rm s}^{-1}$ \\
$C_p(\lam)$ & Planet count rate in the core aperture & e$^-\,{\rm s}^{-1}$ \\
$C_b(\lam)$ & Background count rate & e$^-\,{\rm s}^{-1}$ \\
$g_{\rm geom}(\lam)$ & Geometry factor converting NI in the dark hole to the core aperture & dimensionless \\
$\FRN$ & Flux-ratio noise & flux ratio or ppt \\
$\FRN_{\rm req}$ & Required flux-ratio noise at the design point & flux ratio or ppt \\
$\kappa_{\FRN}$ & Count-rate to FRN conversion factor & flux ratio per e$^-\,{\rm s}^{-1}$ \\
$\kappac$ & Contrast-stability to FRN conversion factor & dimensionless \\
\bottomrule
\end{tabular}
\end{table*}

\begin{table*}[t]
\caption{Core coronagraph, WFSC, covariance, and post-processing notation used throughout the paper.}
\label{tab:nomenclature_model}
\setlength{\tabcolsep}{4pt}
\begin{tabular}{p{0.14\linewidth} p{0.56\linewidth} p{0.18\linewidth}}
\toprule
Symbol & Meaning & Typical units \\
\midrule
$\E(\bm\rho,\lam,t)$ & Coronagraphic complex field in the science focal plane & $\sqrt{\rm NI}$ \\
$\E_0$ & Coherent bias field after WFSC & $\sqrt{\rm NI}$ \\
$\I(\bm\rho,\lam,t)$ & Normalized intensity (NI) & dimensionless \\
$\Craw$ & Mean raw stellar leakage (raw contrast) & NI \\
$\Ccoh$ & Coherent component of raw leakage & NI \\
$\Cinc$ & Incoherent component of raw leakage & NI \\
$\Cvar$ & Mean coherent fluctuation power & NI \\
$\Cstab$ & RMS differential intensity stability metric & NI \\
$J_k=\partial\E/\partial x_k$ & First-order complex-field Jacobian for disturbance coordinate $x_k$ & $\sqrt{\rm NI}/{\rm unit}(x_k)$ \\
$\Gamma_{k\ell}$ & Second-order field sensitivity tensor & $\frac{\sqrt{\rm NI}}{{\rm unit}(x_k){\rm unit}(x_\ell)}$ \\
$S_k$ & Region-averaged scalar sensitivity $\avg{|J_k|^2}_R$ & ${\rm NI}/{\rm unit}(x_k)^2$ \\
$x_k$ & Disturbance coordinate & problem-dependent \\
$\Delta x_k$ & Differential disturbance between calibrated observations & problem-dependent \\
$\sigma_k$ & Residual RMS of $x_k$ & problem-dependent \\
$M_k,\Delta M_k$ & Visit-level mean and change-in-mean state of $x_k$ & problem-dependent \\
$V_k,\Delta V_k$ & Visit-level variance and change-in-variance state of $x_k$ & problem-dependent \\
$H_k(f)$ & Residual transfer function for coordinate $x_k$ & dimensionless \\
$S_{x_k}(f)$ & One-sided PSD of disturbance coordinate $x_k$ & unit$(x_k)^2\,{\rm Hz}^{-1}$ \\
$\SigmaDx$ & Covariance of differenced disturbance states & unit$(x)^2$ \\
$\SigmaDI$ & Covariance of differenced intensity states & NI$^2$ \\
$\SigmaInc$ & Incoherent contribution to the differenced-intensity covariance & NI$^2$ \\
$\bm M$ & Linearized calibration/post-processing operator & dimensionless operator \\
$\eta_{\rm alg}$ & Algorithm throughput for the planet signal & dimensionless \\
$\etapp$ & Residual-speckle transmission of the adopted post-processing pipeline & dimensionless \\
$\Mpp$ & Design margin on speckle uncertainty; not a measured noise factor & dimensionless \\
$\mathcal R_{\rm COPP}$ & Stability-relaxation figure of merit for alternate COPP concepts & dimensionless \\
$\Sigma_d,\Sigma_y$ & Unprocessed and processed data covariance matrices & problem-dependent \\
$\Mmode,\Dmode$ & Coronagraph/WFSC/post-processing mode indices and observing differential indices & --- \\
\bottomrule
\end{tabular}
\end{table*}

\section{Science-facing metric: flux-ratio noise}
\label{sec:frn}

\subsection{Definition}\label{sec:definition}

The appropriate starting point is the quantity inferred from the observation. Expressing uncertainty in planet-to-star flux ratio allows optical, detector, and calibration errors to be compared after their different physical responses have been accounted for. We define flux ratio $f_p(\lam)$ as the planet-to-star flux ratio in a spectral channel centered on $\lam$,
at the telescope entrance pupil, before instrument attenuation. The calibrated estimator corrects for the declared throughput and response; that correction does not redefine the astrophysical flux ratio.
For a required design-point detection signal-to-noise ratio (SNR), SNR$_{\rm req}$ (Sec.~\ref{sec:false_alarm_and_margin}), a convenient top-level requirement is
\begin{equation}
\FRN_{\rm req}(\lam) \equiv \frac{f_p(\lam)}{\mathrm{SNR}_{\rm req}}.
\label{eq:frn_requirement}
\end{equation}
FRN denotes the sampling standard deviation of that calibrated estimator. When its component errors have zero cross covariance, the total is decomposed as
\begin{equation}
\FRN_{\rm tot}^2 = \FRN_{\rm rand}^2 + \FRN_{\rm cal}^2 + \FRN_{\rm speck}^2,
\label{eq:frn_variance_components}
\end{equation}
where $\FRN_{\rm rand}$ is random (photon + detector) noise, $\FRN_{\rm cal}$ is calibration error, and
$\FRN_{\rm speck}$ is the fluctuating residual-starlight contribution \cite{Nemati2020MethodPerformanceSpec,Nemati2023RomanErrorBudget}. Cross-correlated components require $2\sum_{i<j}\Cov(e_i,e_j)$ in Eq.~\eqref{eq:frn_variance_components}. A deterministic error $b_f=\Eavg\widehat f_p-f_p$ is reported separately, with $\mathrm{RMSE}^2=\FRN^2+b_f^2$; it is not independent noise.

\subsection{Planet signal model, phase function, and IWA-limited observability}
\label{sec:planet_signal_geometry}

A useful flux requirement must refer to a planet that is both bright enough and spatially accessible. Reflected-light brightness increases toward full phase, but projected separation decreases, bringing the signal closer to the region rejected by the coronagraph. The phase law and inner working angle (IWA) therefore enter the same allocation problem.

\begin{figure}[t]
\centering
\includegraphics[width=0.52\linewidth]{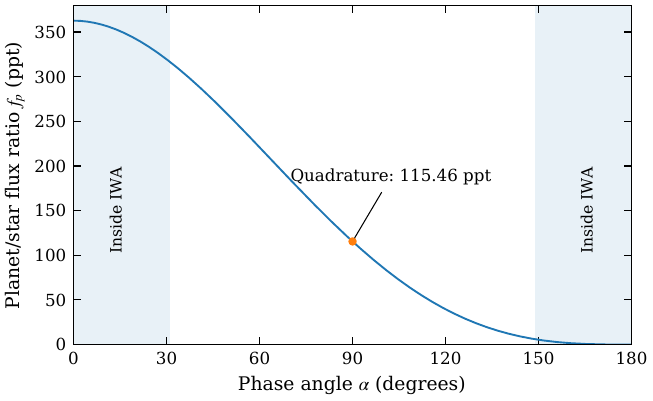}
\caption{Lambertian phase and corresponding Earth-analog flux ratio for $A_g=0.2$, $R_p=6371$ km, and $a=1$ au. The ordinate is the entrance-pupil flux ratio; $\Phi$ is proportional to it under these assumptions. The quadrature value is 115.46~ppt. Shaded intervals fail a monochromatic $3\lambda/D$ IWA at $D=6$ m, $d=10$ pc, and $\lambda=500$ nm: $\alpha<31.04^\circ$ or $\alpha>148.96^\circ$. Geometric access alone does not ensure sufficient throughput or SNR. The count-derived channel example in Sec.~\ref{sec:closure} uses $d=5$ pc so that the retained NIR bands also pass this geometric test.}
\label{fig:phase_fluxratio}
\end{figure}

\paragraph*{Flux ratio versus phase angle.}
For reflected light at wavelength $\lam$, a standard geometric model is
\begin{equation}
f_p(\lam,\alpha) \;=\; A_g(\lam)\left(\frac{R_p}{a}\right)^2 \Phi(\alpha),
\label{eq:fp_phase}
\end{equation}
where $A_g(\lam)$ is the geometric albedo, $R_p$ the planet radius, $a$ the star--planet separation
(assume circular orbit for budgeting), and $\alpha$ is the star--planet--observer phase angle
($\alpha=0$ full phase, $\alpha=\pi/2$ quadrature).
For a Lambertian sphere,
\begin{equation}
\Phi(\alpha) \;=\; \frac{\sin\alpha + (\pi-\alpha)\cos\alpha}{\pi}
\qquad \Rightarrow\qquad
\Phi(\pi/2)=\frac{1}{\pi}.
\label{eq:lambert_phase}
\end{equation}
Figure~\ref{fig:phase_fluxratio} illustrates this competition: the brightest portions of the phase curve can lie inside the IWA, so brightness alone does not identify the most useful observing phase.
For an Earth analog at 1 au, $(R_\oplus/a)^2\simeq1.814\times10^{-9}$ gives a quadrature flux ratio $f_p\sim1.1\times10^{-10}$ at $A_g\simeq0.2$.

\paragraph*{Projected separation and IWA constraint.}
For a circular orbit, the projected star--planet angular separation (working angle) is
\begin{equation}
\mathrm{WA}(\alpha) \;\simeq\; \frac{a}{d}\sin\alpha,
\label{eq:WA_alpha}
\end{equation}
where $d$ is the distance to the system.
In coronagraph coordinates $\rho \equiv \mathrm{WA}/(\lam/\Dtel)$, the IWA constraint is
$\rho(\alpha)\ge \IWA$ where $\IWA$ is in units of $\lam/\Dtel$.
For an accessible circular orbit the allowed phase interval is $\alpha\in[\alpha_{\min},\pi-\alpha_{\min}]$, intersected with the phases reached by its inclination and the visibility/OWA limits, where
\begin{equation}
\alpha_{\min}(\lam,d) \;=\; \arcsin\!\left(\IWA\,\frac{\lam}{\Dtel}\,\frac{d}{a}\right),
\label{eq:alpha_min}
\end{equation}
and if the argument exceeds unity the planet is not observable at that $(\lam,d,a)$ regardless of
integration time (pure geometry). Equivalently, the maximum distance for access at any phase is
\begin{equation}
d_{\max}(\lam) \;=\; \frac{a\,\Dtel}{\IWA\,\lam}.
\label{eq:dmax}
\end{equation}

\paragraph*{Science implication for an error budget.}
Because both $f_p(\alpha)$ and the coronagraph throughput $\tau_{\rm core}(\lam,\rho)$ depend on separation,
the relevant ``design-point'' signal is sometimes screened using the detectability proxy
\begin{equation}
\mathcal{S}(\alpha;\lam)\;\equiv\; f_p(\lam,\alpha)\,\tau_{\rm core}(\lam,\rho(\alpha)),
\label{eq:detectability_proxy}
\end{equation}
subject to $\rho(\alpha)\ge \IWA$ and any $\OWA$ constraint.
Accordingly, the error budget must provide $\Cstab(\lam;\rho)$ (or the covariance it implies) as a
function of working angle, not only annulus averages, to support yield and exposure-time trades \cite{Stark2019YieldLandscape,Savransky2016EXOSIMS}. Equation~\eqref{eq:detectability_proxy} is not an SNR objective when background or residual covariance changes with phase; then the relevant quantity is $f_p^2\bm p^T\bm\Sigma^{-1}\bm p$ for a unit-flux template at the same cost. The Lambert law is a declared phase model, not an atmospheric retrieval: clouds, albedo, radius, orbital geometry, and stellar variability remain astrophysical inputs.

\subsection{Photon-counting model and the conversion factors}
\label{sec:kappa}

For a channel centered at $\lam$ with bandwidth $\Delta\lam$, let $F_\star$ be entrance-pupil stellar photon irradiance (photons s$^{-1}$ m$^{-2}$ nm$^{-1}$), $A_{\rm col}$ collecting area, and $\mathrm{QE}$ detector quantum efficiency. The reference electron rate before coronagraph suppression is
\begin{equation}
C_\star(\lambda_c)=\int_{\Delta\lambda} R_\star(\lambda)\,d\lambda,
\qquad R_\star(\lambda)=A_{\rm col}F_\star(\lambda)T_{\rm ref}(\lambda)\mathrm{QE}(\lambda),
\label{eq:Cstar}
\end{equation}
The common transmission $T_{\rm ref}$ defines the reference for both NI and planet throughput and is applied only once. The planet rate is $C_p=\int R_\star\tau_{\rm core}f_p\,d\lambda$, reducing for constant flux ratio and throughput within the channel to
\begin{equation}
C_p(\lam) = f_p(\lam)\,\tau_{\rm core}(\lam)\,C_\star(\lam).
\label{eq:Cplanet}
\end{equation}

Background means are collected in $C_b(\lam)$ (local zodiacal light, exozodiacal light, and detector dark current), with clock-induced charge accounted for per read. Photon arrivals are conditionally Poisson; structured disks and uncertain background subtraction also require a nuisance/covariance model \cite{Nemati2020MethodPerformanceSpec}.
Residual starlight in the subtraction residual is represented by an \emph{effective} speckle ``rate'' $C_s(\lam)$,
whose interpretation depends on the temporal correlation of residual speckles (Sec.~\ref{sec:correlation}).

\paragraph*{Core-aperture geometry and conversion from dark-hole NI to core counts.}
Let $\Omega_{\rm core}$ be the solid angle of the core aperture and $\Delta\theta$ the detector (or model) pixel scale.
Then $N_{\rm pix,core}=\Omega_{\rm core}/\Delta\theta^2$.
Let $\mathrm{PSF}_{\rm pk}(\lam)$ be the normalized point-spread-function (PSF) peak in units of inverse solid angle.
If $\Cstab(\lam;R)$ is defined as the RMS NI change over a dark-hole region $R$
[Eq.~\eqref{eq:Cstab_def}], then a convenient approximation for the corresponding residual speckle rate in the core aperture is
\begin{equation}
C_s(\lam)\approx g_{\rm geom}(\lam)\;C_\star(\lam)\;\tau_s(\lam)\;\Cstab(\lam;R),
\qquad
g_{\rm geom}(\lam)\equiv \mathrm{PSF}_{\rm pk}(\lam)\,\Omega_{\rm core},
\label{eq:Cs_from_Cstab}
\end{equation}
We use $\tau_s=1$ for detector-plane NI; a nonunit value represents only a separately specified downstream loss absent from the reference and optical model, not the planet throughput. Equation~\eqref{eq:Cs_from_Cstab} assumes a uniform correlated residual in the final planet-aligned aperture. For two-lobe ADI this is the signed residual after registration: identical raw residuals at the two lobes cancel instead. The general pixel-averaged mapping is
\begin{equation}
r_{\star,q}=\int R_\star(\lambda)\,\mathrm{PSF}_{\rm pk}(\lambda)\Omega_q\,\I_q(\lambda)\,d\lambda.
\label{eq:pixel_rate}
\end{equation}
The product $\mathrm{PSF}_{\rm pk}\Omega_q$ is dimensionless. Aperture subdivision conserves the rate and its covariance when the areas and sample covariance transform together. A dark-hole RMS by itself does not determine the variance of a nonuniform extraction.

\paragraph*{Differential-imaging bookkeeping (ADI baseline).}

For the baseline FRN implementation we adopt OS-1 with ADI subtraction, implemented as two equal roll exposures. Let $t$ denote \emph{total} live time and $B=C_{\rm leak}+C_b$ the background rate in each of two disjoint planet apertures. Each signed difference-lobe estimate has variance $2(C_p+2B)/t$ in rate units. Averaging the two independent photon-count estimates gives
\begin{equation}
\Var_{\rm ph}(\widehat C_p)=\frac{V_{\rm ph}}{t},\qquad
V_{\rm ph}=\begin{cases}
 C_p+B,&\text{known mean background, all time on target},\\
 C_p+2B,&\text{two rolls, both disjoint lobes retained},\\
 2(C_p+2B),&\text{equal target and matched-reference times}.
\end{cases}
\label{eq:strategy_photon}
\end{equation}
The first case is an information benchmark, not a free perfect reference. RDI has no second planet-bearing lobe with which to average. These factors apply to photon counting, not automatically to speckle covariance: correlated rolls, overlapping lobes, unequal reference brightness, and planet self-subtraction must be propagated with the actual operator in Sec.~\ref{sec:linop}. For unequal target/reference time fractions $r$ and $1-r$, the photon coefficient is $(C_p+B)/r+B/(1-r)$, minimized at $r=\sqrt{C_p+B}/(\sqrt{C_p+B}+\sqrt B)$. All comparisons charge the reference time and mode-dependent overhead.

\paragraph*{Detector-mediated count-rate model and detector-noise taxonomy.}
\label{sec:detector_model}

Equation~\eqref{eq:Cstar} treats QE as a deterministic response, but the detector chain introduces both random and calibration terms. Use input-photoelectron-referred counts $N_q$ and an output gain $g_q$. Conditional on the optical history, the mean output is
\begin{equation}
\Eavg(n_q\mid\bm x)=g_q[\Lambda_{\gamma,q}+d_qt+n_{{\rm fr},q}c_q]+b_{{\rm elec},q},\qquad
\Lambda_{\gamma,q}=\int \mathrm{QE}_q\,s_q(t')\,dt'.
\label{eq:detector_counts}
\end{equation}
Here $s_q$ is the photon rate before QE, $d_q$ is dark current, $c_q$ is clock-induced charge per read, and $n_{{\rm fr},q}$ is the number of reads. If $s_q$ is already an electron rate, QE is not applied again. A conditional input-referred variance is
\begin{equation}
\Var(N_q\mid\bm x)=F_{\gamma,q}\Lambda_{\gamma,q}+F_{d,q}d_qt+
 n_{{\rm fr},q}(F_{c,q}c_q+r_q^2).
\label{eq:detector_variance}
\end{equation}
The $F$ factors multiply variance, not amplitude. Ideal photon counting has $F_\gamma=1$; analog electron multiplication can approach $F_\gamma=2$, while thresholding requires explicit detection-efficiency and coincidence-loss calibration. Output-referred variance multiplies the bracket by $g_q^2$, with any additional output noise added afterwards. Cosmic-ray masks change samples and live time; a generic additive Gaussian ``cosmic-ray reserve'' does not specify that loss.

For disjoint ideal detector samples with conditional Poisson mean vector $\bm\Lambda$, the law of total covariance gives
\begin{equation}
\Cov(\bm N)=\diag(\Eavg\bm\Lambda)+\Cov(\bm\Lambda).
\label{eq:total_count_cov}
\end{equation}
Equation~\eqref{eq:total_count_cov} is the statistical reason to keep mean leakage and stability separate. Subtraction can remove the expected stellar image, but it cannot undo the random arrival of its photons. Variations of the optical state add correlated fluctuations of the count means, represented by the second term. The two terms describe different randomness and are both required.

The conditional detector variance belongs in $\FRN_{\rm rand}$; uncertainty in QE, gain, dark current, clock-induced charge, or electronic bias belongs in the calibration covariance of Sec.~\ref{sec:signalextraction}. Roman detector and camera-calibration studies provide the relevant engineering heritage \cite{Morrissey2023RomanEMCCD,Bush2025RomanCameraSystems}.

\subsection{Exposure-time and FRN closure}
\label{sec:frn_closure}

\paragraph*{Mean residual starlight photon noise.}

The mean residual stellar rate before subtraction follows from the raw leakage at the planet location,
\begin{equation}
C_{\rm leak}(\lam)\;\approx\; g_{\rm geom}(\lam)\;C_\star(\lam)\;\tau_s(\lam)\;\Craw(\lam;\bm\rho_p),
\label{eq:Cleak}
\end{equation}
with $g_{\rm geom}$ defined in Eq.~\eqref{eq:Cs_from_Cstab}.
In annulus-averaged budgets one may replace $\Craw(\lam;\bm\rho_p)$ by an appropriate regional average such as
$\avg{\I}_{R}$ or $\Ccoh(\lam;R)+\Cvar(\lam;R)+\Cinc(\lam;R)$ (Sec.~\ref{sec:raw_contrast}).

\paragraph*{Fully correlated speckle-systematics limit.}

If the \emph{residual} speckle term (after calibration/subtraction) is treated as fully correlated over the exposure
(conservative limit), the integrated SNR after time $t$ is
\begin{equation}
\mathrm{SNR}(t)=\frac{C_p t}{\sqrt{V_{\rm ph}t+(C_st)^2}}.
\label{eq:SNR_correlated}
\end{equation}
Solving Eq.~\eqref{eq:SNR_correlated} for the time to reach SNR$_{\rm req}$ yields
\begin{equation}
t_{\rm req}=\frac{\mathrm{SNR}_{\rm req}^2 V_{\rm ph}}
{C_p^2-\mathrm{SNR}_{\rm req}^2C_s^2},\qquad
T_{\rm wall}=t_{\rm req}/\eta_{\rm live}.
\label{eq:t_req}
\end{equation}
Here $V_{\rm ph}$ follows Eq.~\eqref{eq:strategy_photon}, additional white detector terms enter the same coefficient, and $\eta_{\rm live}$ is the mode-specific live fraction. The denominator identifies two different design regimes. When the persistent residual is small, additional exposure reduces photon uncertainty; as $\mathrm{SNR}_{\rm req}C_s$ approaches $C_p$, the required time diverges. Once that ceiling is reached, more integration cannot substitute for improved stability or calibration. The expression assumes an unbiased, stationary scene and a zero-mean residual persistent over the program; an unknown fixed bias requires a separate detection test.

\subsection{Flux-ratio-noise conversion factors}
\label{sec:kappa_factors}

For a fixed allocated integration time $t_{\rm alloc}$, it is useful to work directly in FRN.
Define the conversion factor between electron-rate noise and flux-ratio noise as
\begin{equation}
\kappa_{\FRN}(\lambda)=\frac{1}{C_\star(\lambda)\tau_{\rm core}(\lambda)},\qquad
\kappa_N(\lambda)=\frac{\kappa_{\FRN}(\lambda)}{t_{\rm alloc}}.
\label{eq:kappa_def}
\end{equation}
so that a noise term expressed as an electron-rate uncertainty $\sigma_C$ maps to FRN as
$\FRN = \kappa_{\FRN}\,\sigma_C$, whereas a count uncertainty $\sigma_N$ gives $\FRN=\kappa_N\sigma_N$. For a nonconstant spectrum the denominator is the response to the declared unit-flux spectral template.

Similarly, define the factor that converts \emph{contrast stability} (NI RMS over dark-hole pixels) into
flux-ratio noise as
\begin{equation}
\kappac(\lambda)=\frac{\mathrm{PSF}_{\rm pk}(\lambda)\Omega_{\rm core}\tau_s(\lambda)}{\tau_{\rm core}(\lambda)}
=\frac{g_{\rm geom}(\lambda)\tau_s(\lambda)}{\tau_{\rm core}(\lambda)}.
\label{eq:kappa_c_def}
\end{equation}

Then, under the approximation in Eq.~\eqref{eq:Cs_from_Cstab} and with the response normalized back to the entrance-pupil planet flux, the budgeted stochastic FRN contribution from contrast stability is
\begin{equation}
\FRN_{\rm speck,budget}(\lambda)=\Mpp(\lambda)\frac{\etapp(\lambda)}{\eta_{\rm alg}(\lambda)}\kappac(\lambda)\Cstab(\lambda;R).
\label{eq:FRN_speck}
\end{equation}

The measured residual transmission $\etapp\ge0$ and planet throughput $\eta_{\rm alg}$ describe the same pipeline, whereas $\Mpp\ge1$ inflates budgeted uncertainty without being an observed noise statistic. Derive $\etapp$ from the processing operator and estimator, not a universal suppression factor. Planet attenuation enters $C_p\to\eta_{\rm alg}C_p$; when $\kappac$ already uses processed throughput, do not divide by $\eta_{\rm alg}$ again.

\subsection{Partial correlation of speckle residuals}
\label{sec:correlation}

Eq.~\eqref{eq:SNR_correlated} is the conservative limit where the post-subtraction residual behaves as a
fully correlated systematic over the exposure.
A compact and more realistic closure is a two-component model in the measurement aperture:
(i) a quasi-static systematic term with rate $C_{s,{\rm sys}}$ (does not average down), and
(ii) a stochastic term with rate scale $C_{s,{\rm rand}}$ and correlation time $\tau_c$ (partially averages down).
For the explicitly adopted exponential rate covariance $K_s(\Delta t)=C_{s,{\rm rand}}^2e^{-|\Delta t|/\tau_c}$, the total variance after time $t$ is
\begin{equation}
\Var(t)=V_{\rm ph}t+(C_{s,{\rm sys}}t)^2+
2C_{s,{\rm rand}}^2\left[\tau_ct-\tau_c^2(1-e^{-t/\tau_c})\right].
\label{eq:Var_twocomp}
\end{equation}
The stochastic contribution tends to $C_{s,{\rm rand}}^2t^2$ for $t\ll\tau_c$ and $2C_{s,{\rm rand}}^2\tau_ct$ for $t\gg\tau_c$. Setting $C_{s,{\rm sys}}=0$ removes the persistent floor, not the finite correlation time. Actual roll/reference windows are applied in Sec.~\ref{sec:temporal}. In covariance-based implementations (Sec.~\ref{sec:signalextraction}),
$C_{s,{\rm sys}}$ corresponds to low-rank/slow modes and $C_{s,{\rm rand}}$ to broadband components
\cite{Soummer2007SpeckleNoise,GrecoBrandt2016SpectralCovariance}.

The remaining task is to determine these rate statistics from the optical system. The next section separates physical coherence from temporal variability, so that the same residual light is neither omitted from the photon budget nor counted twice as an independent instability.

\section{Optical model: coherent and incoherent raw contrast}
\label{sec:raw_contrast}

\begin{figure}[t]
\centering
\begin{tikzpicture}[font=\small,box/.style={draw,line width=.45pt,align=center,text width=6.0cm,minimum height=1.0cm,inner sep=5pt},arrow/.style={-{Stealth[length=1.6mm]},line width=.45pt}]
\node[box,text width=12.8cm] (fields) at (0,0) {Physical coherence channels and shared disturbances\\$I_q=\sum_s w_s|E_{0,qs}+J_{qs}x+\cdots|^2+I_{{\rm ext},q}$};
\node[box] (mean) at (-3.45,-1.8) {Mean stellar leakage\\$C_{\rm raw}=C_{\rm coh}+C_{\rm var}+C_{\rm inc}$};
\node[box] (difference) at (3.45,-1.8) {Calibrated observing differences\\Differential mean $\mu_{\Delta I}$ and covariance $\Sigma_{\Delta I}$};
\node[box] (photon) at (-3.45,-3.6) {Unsubtracted photon means $\Lambda$\\Shot noise: $\operatorname{diag}(\mathbb E\Lambda)$};
\node[box] (optical) at (3.45,-3.6) {Optical statistics in rate/count units\\$\operatorname{Cov}(\Lambda)$; bias retained separately};
\node[box,text width=12.8cm] (estimator) at (0,-5.5) {Detector response, calibration, and the actual processing/extraction operator\\Planet template $p$; flux bias $b_f$; $\mathrm{FRN}^2=w^T\Sigma_yw$; $\mathrm{RMSE}^2=\mathrm{FRN}^2+b_f^2$};
\draw[arrow] (fields.south) -- ++(0,-.35) -| (mean.north);
\draw[arrow] (fields.south) -- ++(0,-.35) -| (difference.north);
\draw[arrow] (mean) -- (photon);
\draw[arrow] (difference) -- (optical);
\draw[arrow] (photon.south) -- ++(0,-.4) -| (estimator.north);
\draw[arrow] (optical.south) -- ++(0,-.4) -| (estimator.north);
\node[box,text width=12.8cm,minimum height=1.05cm] (accept) at (0,-7.25) {Allocation acceptance at the same target, estimator, and observing cost\\Calibrated null/planet-present laws: $P_{\rm FA,tot}\le\alpha_{\rm tot}$ and $P_{\rm D}\ge1-\beta$};
\draw[arrow] (estimator.south) -- (accept.north);
\end{tikzpicture}
\caption{Bookkeeping from optical fields to the science estimator. Every mutually incoherent optical channel can have its own coherent bias--perturbation interference. Temporal fluctuation power contributes to the mean but is not a new physical incoherent source. All channels supply mean photon counts; their intensity variations and shared disturbances can supply cross covariance. Bias is retained separately from the covariance and from FRN. The final probability test additionally requires a calibrated disturbance/detector law: moments alone do not specify its tails. Arrows describe propagation and acceptance, not independent RSS categories.}
\label{fig:contrast_tree}
\end{figure}
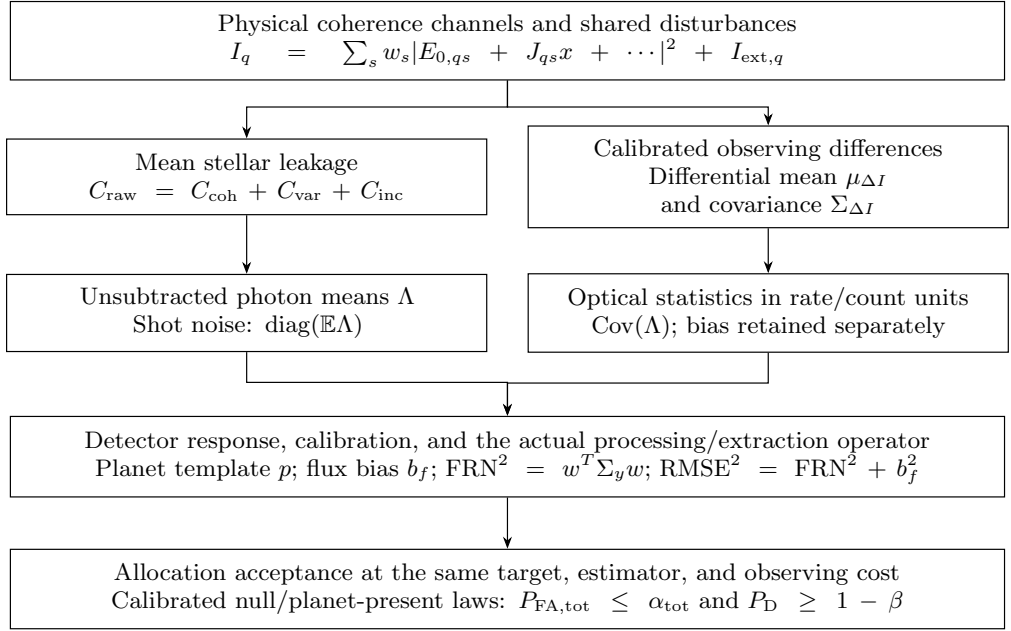

The optical decomposition serves two purposes: it determines which fields interfere and which intensity statistics enter the measurement. These are not the same classification. Figure~\ref{fig:contrast_tree} follows the resulting paths to photon noise, residual covariance, and the final detection test.

\subsection{Normalization and averaging conventions}\label{sec:normalization_and_averaging_conventions}
We define the normalized intensity (NI) such that a non-coronagraphic on-axis stellar PSF has unit peak:
$I_{\star,{\rm pk}}(\lam)\equiv 1$. The coronagraphic NI in the science image is then
\begin{equation}
\I(\bm\rho,\lam,t)\equiv \frac{I_{\rm det}(\bm\rho,\lam,t)}{I_{\star,{\rm pk}}(\lam)} = |\E(\bm\rho,\lam,t)|^2,
\label{eq:normalized_intensity}
\end{equation}
NI is dimensionless, but is not automatically the off-axis-throughput-corrected planet-to-star contrast. Liu et al. normalize sensitivities by the off-axis PSF peak at the planet \cite{Liu2026EAC}. If $r_{\rm pk}=I_{\rm pk,off}/I_{\star,\rm pk}$ in the same optical reference, then $I_{\rm EAC}=I_{\rm NI}/r_{\rm pk}$ and $J_{\rm EAC}=J_{\rm NI}/\sqrt{r_{\rm pk}}$ for fixed normalization. This peak factor is generally not the aperture throughput $\tau_{\rm core}$. Project allocations require this conversion and the same extraction covariance before comparison. The reference PSF density integrates to unity over solid angle, and its peak is $\mathrm{PSF}_{\rm pk}$ used in Eq.~\eqref{eq:pixel_rate}.
For any quantity $Q(\bm\rho)$ and region $R$ (set of pixels or a continuous domain), we use the explicit region average
\begin{equation}
\avg{Q}_{R}\;\equiv\;\frac{1}{\Omega_R}\int_{R} Q(\bm\rho)\,d\Omega
\qquad\text{(or } \avg{Q}_{R}\equiv \frac{1}{N_R}\sum_{\bm\rho_q\in R} Q_q \text{ in discrete form)}.
\label{eq:region_average}
\end{equation}

\subsection{Coherent vs.\ incoherent starlight channels}\label{sec:coherent_vs_incoherent_starlight_channels}
We resolve the stellar radiation into physically mutually incoherent channels (e.g., independent source directions or orthogonal input-polarization states). With weights $w_m\ge0$ normalized to the same total input flux, $\sum_m w_m=1$, the detected NI is
\begin{equation}
\I=\sum_m w_m|\E_m|^2+I_{\rm ext},\qquad \E_m=\E_{0,m}+\delta\E_m.
\label{eq:incoherent_channels}
\end{equation}
The fields of mutually incoherent channels do not interfere, but each perturbation still interferes with its own channel's bias field. Moreover, a common mirror or pointing motion can correlate the channels' intensities. Physical incoherence therefore determines how fields are added; it does not establish statistical independence of their residuals. A scatter or ghost path is assigned to $I_{\rm ext}$ only when bandwidth, path delay, polarization, or source averaging destroys its mutual coherence with the retained fields.

\subsection{Mean leakage vs.\ stability-relevant leakage}\label{sec:mean_leakage_vs_stability_relevant_leakage}
Let $\Ecoh=\E_0+\delta\E$ for a single dominant channel, with $\E_0=\Eavg_t\Ecoh$ so that $\Eavg_t\delta\E=0$. Then
\begin{equation}
\Craw=\underbrace{\avg{|\E_0|^2}_R}_{\Ccoh}
+\underbrace{\avg{\Eavg_t|\delta\E|^2}_R}_{\Cvar}
+\underbrace{\avg{I_{\rm inc}}_R}_{\Cinc}.
\label{eq:mean_raw_split}
\end{equation}
If $\E_0$ is a nominal rather than a mean field, add $2\avg{\Ree(\E_0^*\Eavg_t\delta\E)}_R$. For multiple channels, sum the corresponding terms with their flux weights. The positive fluctuation power $\Cvar$ contributes photon noise whether it is fast or slow; it is not physically incoherent solely because its first moment vanishes. Both linear mixing and changes in quadratic power can survive a finite observing differential (Secs.~\ref{sec:stability}--\ref{sec:temporal}). Figure~\ref{fig:contrast_tree} preserves these distinct paths through the budget.

\paragraph*{Diffuse narrow-angle scatter as a separate incoherent branch.}
Surface micro-roughness, particulate contamination, and coating-induced diffuse narrow-angle scatter can supply an intensity-only halo when the coherence test above is met; otherwise they remain in the coherent propagation. A compact bookkeeping closure is
\begin{equation}
C_{\rm raw}(\lambda;\rho)\rightarrow C_{\rm raw}(\lambda;\rho)+C_{\rm NAS}(\lambda;\rho),
\qquad
C_{\rm inc,stat}(\lambda;\rho)\rightarrow C_{\rm inc,stat}(\lambda;\rho)+C_{\rm NAS}(\lambda;\rho),
\label{eq:scatter_mean_budget}
\end{equation}
where $C_{\rm NAS}(\lambda;\rho)$ is the mean narrow-angle scatter halo in normalized-intensity units.
The corresponding contribution to the mean leakage rate in the measurement aperture is
\begin{equation}
C_{\rm leak}^{\rm mean}(\lambda)\rightarrow
C_{\rm leak}^{\rm mean}(\lambda)
+
g_{\rm geom}(\lambda)\,C_\star(\lambda)\,\tau_s(\lambda)\,C_{\rm NAS}(\lambda;\rho_p).
\label{eq:scatter_count_rate}
\end{equation}

If the separately modeled scatter halo varies between calibrated observations and satisfies $\Eavg\avg{\Delta I_{\rm other}\Delta C_{\rm NAS}}_R=0$, its mean-square contribution may be carried as
\begin{equation}
C_{\rm stab}^2 \rightarrow C_{\rm stab}^2 + \sigma_{\rm NAS}^2,
\qquad
\sigma_{\rm NAS}^2 \equiv \Eavg\avg{\left(\Delta C_{\rm NAS}\right)^2}_{R}.
\label{eq:scatter_difference}
\end{equation}
Independence with zero means is sufficient for this condition; uncorrelated fluctuations with nonzero means are not. Otherwise add the cross term $2\Eavg\avg{\Delta I_{\rm other}\Delta C_{\rm NAS}}_R$. This separation is useful when flowing the error budget to mirror microstructure, cleanliness, and coating requirements, where scatter may be a dominant mirror-level specification even when it does not participate in
coherent mixing. \cite{NematiStahl2025ScatterFlowdown}

\subsection{Static incoherent leakage and the photometric branch}
\label{sec:photometric_branch}

For the mean photon budget, collect the static or slowly varying intensity-only channels as
\begin{equation}
\Cincstat=C_{\rm pol,inc}+C_{\rm NAS,inc}+C_{\rm ghost,inc}+C_{\rm other,inc}.
\label{eq:Cincstat}
\end{equation}
The terms denote physically incoherent, intensity-only channels evaluated at the planet or over a specified region. Pointing, WFE, and beamwalk variance instead enter $\Cvar$ through the coherent-channel covariance. Polarization labels include only the part that passes the channel coherence accounting of Sec.~\ref{sec:pol}.

The corresponding mean stellar leakage rate in the measurement aperture is
\begin{equation}
C_{\rm leak}^{\rm mean}=g_{\rm geom}C_\star\tau_s[\Ccoh+\Cvar+\Cincstat].
\label{eq:Cleak_mean_split}
\end{equation}
Its photon-noise contribution to $\FRN_{\rm rand}$ is
\begin{equation}
\FRN_{\rm leak,shot}(\lambda)
=
\frac{\sqrt{C_{\rm leak}^{\rm mean}(\lambda)\,t_{\rm alloc}}}
{C_\star(\lambda)\,\tau_{\rm core}(\lambda)\,t_{\rm alloc}}.
\label{eq:FRN_leak_shot}
\end{equation}
This single-exposure contribution acquires the observing weights of Eq.~\eqref{eq:strategy_photon} in differential measurements.

For variance-driven disturbances (tip--tilt jitter, beamwalk, or LOWFS residuals), define $v_k\equiv\Var[x_k]$ and expand the mean fluctuation leakage in those coordinates:
\begin{equation}
\Delta C_{\rm var}^{(k)}(\lambda;R)
\approx
\left.\frac{\partial \avg{\I}_{R}}{\partial v_k}\right|_0 \Delta v_k
+
\frac{1}{2}\sum_{\ell}
\left.\frac{\partial^2 \avg{\I}_{R}}{\partial v_k\,\partial v_\ell}\right|_0
\Delta v_k\,\Delta v_\ell .
\label{eq:Cinc_variance_states}
\end{equation}
Mean jitter power and its inter-visit change therefore produce distinct photon and differential-stability requirements.

When a compact scalar allocation is desired, one may write
\begin{equation}
\Cphot(\lambda;\rho)\equiv\Cvar+\Cincstat=\sum_i C_{{\rm phot},i}(\lambda;\rho).
\label{eq:Cphot_alloc}
\end{equation}
The labels $i$ describe nonoverlapping positive mean-intensity contributions. Correlated coherent modes must first be combined in $\Eavg|\sum_kJ_kx_k|^2$, not assigned independent positive diagonal terms. Their mean leakage and its uncertainty are different budget quantities.

An exozodiacal comparison uses $C_{\rm zodi,eq}=C_{\rm zodi}/(g_{\rm geom}C_\star\tau_s)$; uncertain disk structure remains a nuisance term. Reducing a stable halo improves photon statistics, whereas reducing its poorly calibrated changes improves differential stability. The field response below determines both effects.

\section{Linearized disturbance-to-speckle model}
\label{sec:linear}

\subsection{Complex-field Jacobians}\label{sec:complex_field_jacobians}

Let $x_k(t)$ be a set of disturbance coordinates (Zernike coefficients, segment rigid-body modes,
beamwalk coordinates, DM drift states, etc.). We define the complex-field Jacobian maps
\begin{equation}
J_k(\bm\rho,\lam)\equiv \frac{\partial \E(\bm\rho,\lam)}{\partial x_k},
\label{eq:field_jacobian}
\end{equation}
evaluated about the bias state after high-order WFSC.
For small residual disturbances $\delta x_k(t)$, the field perturbation is
\begin{equation}
\Delta \E(\bm\rho,\lam,t)=\sum_k J_k(\bm\rho,\lam)\,\delta x_k(t).
\label{eq:DeltaE}
\end{equation}
This linearization is the same object exploited by electric-field-based wavefront sensing and control
(e.g., electric field conjugation (EFC) combined with pairwise probing), and is naturally produced by integrated modeling tools \cite{Giveon2007EFC,Riggs2018FALCO4,Krist2023ModelingRoman}.

\subsection{Intensity expansion and the coherent mixing term}\label{sec:intensity_expansion_and_the_coherent_mixing_term}

Let $\E_0(\bm\rho,\lam)$ be the initial coherent dark-hole field for a given calibrated state.
Then the instantaneous intensity with perturbations is
\begin{align}
|\E_0+\Delta \E|^2
&=|\E_0|^2 + 2\Ree\{\E_0\Delta\E^\ast\} + |\Delta \E|^2.
\label{eq:I_expand}
\end{align}
The residual stellar field acts as a coherent reference for the perturbation: a small displacement changes how two amplitudes interfere, rather than merely adding the perturbation's own light. The cross term is therefore linear in displacement, whereas $|\Delta\E|^2$ is quadratic. This is stellar self-interference, not interference between star and planet. It explains why a faint mean halo can remain sensitive to very small mechanical or thermal changes. The response is largest where the bias and perturbation overlap in phase, and can vanish where their phases are in quadrature.

\paragraph*{Relative scaling of the two contributions.}
Let $|E_0|^2\sim \Ccoh$ and let a single residual mode satisfy $\avg{|J_k|^2}_{R}\sim S_k$.
If $\delta x_k$ is zero-mean with RMS $\sigma_k$, the cross-term RMS scales as
\begin{equation}
\sigma_{\Delta I,{\rm cross}}\;\sim\;2\,\sqrt{\Ccoh\,S_k}\;\sigma_k,
\label{eq:mixing_scale}
\end{equation}
whereas the quadratic term scales as $\sigma_{\Delta I,{\rm quad}}\sim S_k\,\sigma_k^2$.
At fixed nonzero phase and spatial overlap, a deeper coherent dark hole reduces the linear response as $\sqrt{\Ccoh}$, whereas reducing the disturbance suppresses its own intensity as $\sigma_k^2$. These are different design levers. At a phase or overlap null, however, the linear term vanishes: $\sigma_k\ll2\sqrt{\Ccoh/S_k}$ alone does not establish its dominance. Section~\ref{sec:piston_example} quantifies the phase and overlap dependence rather than assigning a universal picometer requirement.

\subsection{Region-averaged sensitivity scalars}\label{sec:region_averaged_sensitivity_scalars}

For region $R$ and wavelength $\lam$, define scalar sensitivities
\begin{equation}
S_k(\lam;R)\equiv \avg{|J_k|^2}_{R}.
\label{eq:jacobian_power}
\end{equation}
For centered, independent coordinates, the mean quadratic-field power is
\begin{equation}
\Eavg\avg{|\Delta\E|^2}_R=\sum_k S_k\sigma_k^2.
\label{eq:quadratic_power}
\end{equation}
For general covariance $\bm V$ it is $\sum_{k\ell}\avg{J_kJ_\ell^*}_R V_{k\ell}$, with $|\bm J\bm\mu|^2$ added for nonzero means. Independence of coordinates does not specify the bias--Jacobian field phase. Define the exact single-coordinate linear-response magnitude
\begin{equation}
G_k^2\equiv\avg{[2\Ree(\E_0^*J_k)]^2}_R
=4\Ccoh S_k\,\mathcal O_k\mathcal P_k,
\label{eq:phase_response}
\end{equation}
where $\mathcal O_k=\avg{|\E_0|^2|J_k|^2}_R/(\Ccoh S_k)$ measures spatial overlap and
$\mathcal P_k=\avg{|\E_0|^2|J_k|^2\cos^2\theta_k}_R/\avg{|\E_0|^2|J_k|^2}_R$, with $\theta_k=\arg J_k-\arg\E_0$, measures phase. If the overlap vanishes, $G_k=0$ directly. Only $\mathcal P_k$ is restricted to $[0,1]$; $\mathcal O_k$ need not be at most one.
The magnitude-only surrogate
\begin{equation}
\sigma_{\Delta I,\rm cross}(R)\simeq\sqrt{2\Ccoh\sum_k S_k\sigma_k^2}
\label{eq:sigma_cross}
\end{equation}
additionally assumes $\mathcal O_k\simeq1$ and $\mathcal P_k\simeq1/2$. This factorization explains what is lost by retaining only field magnitudes: the bias and perturbation may occupy different parts of the dark hole, or overlap while interfering weakly. Neither behavior follows from their separate regional powers. Complex maps are needed to distinguish a genuinely weak response from an unjustified phase average.

\subsection{Second-order field model and nonlinearity closure}
\label{sec:second_order_field}

The first-order Jacobian model is sufficient only if second-order field terms remain subdominant over the
residual disturbance envelope.
We therefore write the field perturbation as
\begin{equation}
\Delta \E(\bm\rho,\lambda,t)
=
\sum_k J_k(\bm\rho,\lambda)\,\delta x_k(t)
+
\frac{1}{2}\sum_{k,\ell}
\Gamma_{k\ell}(\bm\rho,\lambda)\,
\delta x_k(t)\,\delta x_\ell(t)
+
\mathcal{O}(\delta x^3),
\label{eq:DeltaE_second_order}
\end{equation}
where $\Gamma_{k\ell}(\bm\rho,\lambda)\equiv\partial^2\E(\bm\rho,\lambda)/(\partial x_k\,\partial x_\ell)$ is the second-order field sensitivity.

The ratio of second- to first-order field norms is a screening diagnostic, not a bound on a projected intensity covariance. At an overlap or phase null, the coherent-mixing term can vanish even for arbitrarily small perturbations. Acceptance therefore compares bias and estimator covariance with nonlinear propagation over the intended disturbance distribution (Sec.~\ref{sec:validation}).

For a Gaussian disturbance vector $\bm x\sim\mathcal N(\bm\mu,\bm V)$, a complete second-order \emph{intensity} model is
\begin{align}
I_q&=c_q+\bm a_q^T\bm x+\bm x^T\bm Q_q\bm x,\qquad
 a_{qk}=2\Ree(E_{0,q}^*J_{qk}),\nonumber\\
Q_{q,k\ell}&=\Ree(J_{qk}^*J_{q\ell}+E_{0,q}^*\Gamma_{q,k\ell}),
\label{eq:quadratic_intensity}\\
m_q&=c_q+\bm a_q^T\bm\mu+\bm\mu^T\bm Q_q\bm\mu+\Tr(\bm Q_q\bm V),\nonumber\\
[\bm\Sigma_I]_{qr}&=\bm h_q^T\bm V\bm h_r+2\Tr(\bm Q_q\bm V\bm Q_r\bm V),\quad
\bm h_q=\bm a_q+2\bm Q_q\bm\mu.
\label{eq:gaussian_moments}
\end{align}
For incoherent channels, sum their $c$, $\bm a$, and $\bm Q$ with the flux weights before taking moments: the same $\bm x$ can correlate their intensities. The first covariance term describes coherent mixing about the actual disturbance mean; the trace term describes fluctuations of quadratic intensity. The latter remains even when the projected linear response vanishes, and changes in $\bm V$ can change the mean leakage through $\Tr(\bm Q_q\bm V)$. Equation~\eqref{eq:gaussian_moments} thus separates physical effects that a single WFE RMS would combine. It is an exact Gaussian polynomial identity, not an assumption that intensity is Gaussian. Appendix~\ref{app:moments} gives the derivation and the joint-visit construction. It is exact for an exactly linear optical field, but for general optics Eq.~\eqref{eq:quadratic_intensity} is a truncation. At $J_q=0$, the quartic intensity $|\bm x^T\bm\Gamma_q\bm x/2|^2$ may be leading; retaining a squared quadratic field then requires moments through eighth order and comparison with the omitted higher field orders. Small global field error alone does not establish local relative accuracy.

\section{Contrast stability metric and allocation flowdown}
\label{sec:stability}

\subsection{Definition of contrast stability for differential imaging}\label{sec:definition_of_contrast_stability_for_differential_imaging}

Consider a differential measurement between two calibrated observations, $A$ and $B$. In the baseline OS-1 branch these are the two roll-differenced ADI states; in alternate modes they may represent a target/reference pair or another calibrated differential. Define
\begin{equation}
\Delta \I(\bm\rho,\lam)\equiv \avg{\I_A(\bm\rho,\lam,t)}_{t\in A} -
\avg{\I_B(\bm\rho,\lam,t)}_{t\in B}.
\label{eq:visit_intensity_difference}
\end{equation}
The stability metric adopted in several HWO and Roman/CGI budgeting studies is the root-mean-square (RMS) of this difference over $R$,
\begin{equation}
\Cstab(\lambda;R)\equiv\left[\Eavg\avg{(\Delta I)^2}_R\right]^{1/2}.
\label{eq:Cstab_def}
\end{equation}
The expectation refers to repeated realizations of the specified observing sequence; a measured spatial RMS is one realization. This metric is useful for optical allocation, but it does not yet specify how a residual pattern projects onto the planet estimator. The following covariance form retains that information for the extraction in Sec.~\ref{sec:signalextraction}.

\subsection{Vector form, covariance, and category-level RSS}\label{sec:vector_form_covariance_and_category_level_rss}
Let $\Delta\bm I$ denote the stacked NI difference over all samples in $R$ (pixels and, if desired, wavelengths).
Under coherent-mixing dominance, the linearized differenced model can be written as
\begin{equation}
\Delta\bm I \;\simeq\; \bm A\,\Delta\bm x,
\qquad
A_{qk}(\lambda)\equiv 2\,\Re\!\left\{E_0(\bm\rho_q,\lam_q)\,J_k^\ast(\bm\rho_q,\lam_q)\right\}.
\label{eq:A_matrix}
\end{equation}
Then the ensemble covariance in this linear limit is
\begin{equation}
\bm\Sigma_{\Delta I}(\lambda)\;=\;\bm A\,\bm\Sigma_{\Delta x}\,\bm A^{\mathsf T}(\lambda),
\label{eq:Sigma_DI}
\end{equation}
and, for equal-area samples, the stability metric is
\begin{equation}
\Cstab^2=\frac{\Tr\bm\Sigma_{\Delta I}+\bm\mu_{\Delta I}^T\bm\mu_{\Delta I}}{N_R},\qquad
\bm\mu_{\Delta I}=\Eavg\Delta\bm I.
\label{eq:Cstab_trace}
\end{equation}
For unequal areas replace $1/N_R$ by normalized area weights. A repeatable differential bias can have zero covariance and nonzero $\Cstab$. Equation~\eqref{eq:FRN_speck} applies to the zero-mean fluctuating part; otherwise its NI input is $[\Tr(\bm W_R\bm\Sigma_{\Delta I})]^{1/2}$, while $\bm\mu_{\Delta I}$ must be propagated separately into estimator bias. For unequal optical states, nonzero disturbance means, or significant quadratic leakage, apply Eq.~\eqref{eq:gaussian_moments} to the joint state $(\bm x_A^T,\bm x_B^T)^T$ as specified in Appendix~\ref{app:moments}.

A category-level stability budget is commonly expressed as a root-sum-square (RSS) allocation,
$\Cstab^2=\sum_{i=1}^{N_{\rm cat}} C_{{\rm stab},i}^2$, but this is valid when category residuals have zero cross second moments in the differenced observable [Eq.~\eqref{eq:Cstab_def}]. For centered categories, zero cross covariance suffices; independence alone does not remove products of nonzero means.
A practical implementation defines categories with their cross covariance, rather than treating a model-uncertainty factor (MUF) as a substitute for independence. If the $i$th category has zero-mean projected standard deviation $s_i$, its total variance is $\sum_i s_i^2+2\sum_{i<j}\Cov(e_i,e_j)$. With unknown correlations the conservative bound is $(\sum_i s_i)^2$, not an unqualified RSS. Deterministic category biases add with their signs, or with absolute-value bounds for a worst-case allocation. Equations~\eqref{eq:A_matrix}--\eqref{eq:Cstab_trace} expose the structural-mode couplings that an RSS budget otherwise hides \cite{Nemati2023RomanErrorBudget}.

\subsection{Single-disturbance mapping to allowable residual amplitude}\label{sec:single_disturbance_mapping_to_allowable_residual_amplitude}

Assume the dominant stability contribution is the coherent mixing term and that the differential residual between
two calibrated observations is $\Delta x_k = x_{k,A}-x_{k,B}$.
Rather than assuming independence a priori, define
\begin{equation}
\Var(\Delta x_k)=2\sigma_k^2(1-\rho_k),
\label{eq:visit_difference_variance}
\end{equation}
where $\sigma_k^2\equiv\Var(x_k)$ is the single-observation residual variance and
$\rho_k\equiv {\rm Corr}(x_{k,A},x_{k,B})$ is the correlation coefficient between the averaged residual states in the two
observations. The independent case corresponds to $\rho_k=0$, while quasi-static common-mode behavior gives
$\rho_k\rightarrow 1$.

For equal single-visit variances, zero differential mean, and a common field/Jacobian, Eq.~\eqref{eq:phase_response} gives
\begin{equation}
C_{{\rm stab},k}=G_k\sqrt{\Var(\Delta x_k)}
=G_k\sqrt{2(1-\rho_k)}\,\sigma_k.
\label{eq:Cstab_single}
\end{equation}
Whenever a temporal model is available, $\Var(\Delta x_k)$ should be computed from Eq.~\eqref{eq:PSD_diff} rather than treated as a free parameter. Solving for the allowable residual disturbance RMS:
\begin{equation}
\sigma_k\le\frac{C_{{\rm stab},k,\rm alloc}}{\mathrm{MUF}_k\,G_k\sqrt{2(1-\rho_k)}}.
\label{eq:sigma_alloc}
\end{equation}
Here $\mathrm{MUF}_k\ge1$ bounds uncertainty in the \emph{specified} response, e.g. $G_{k,\rm true}\le\mathrm{MUF}_kG_{k,\rm model}$ over a declared parameter envelope. It is not a confidence level, nor a bound on an omitted response direction. Equation~\eqref{eq:sigma_alloc} is inapplicable at $G_k=0$ without a higher-order model. For unequal state variances use $\Var(\Delta x_k)=V_{A,kk}+V_{B,kk}-2V_{AB,kk}$; a deterministic change in mean consumes a separate bias allocation.

\subsection{Disturbance coordinates, units, and mapping to integrated-model Jacobians}\label{sec:disturbance_coordinates_units_and_mapping_to_integrated_model_jacobians}

The disturbance vector $\bm x$ is a bookkeeping device: it must be chosen so that (i) the integrated model can
provide $J_k=\partial E/\partial x_k$, and (ii) the observatory can provide a disturbance PSD and control rejection
for each $x_k$ (Sec.~\ref{sec:temporal}). Table~\ref{tab:disturbance_coords} summarizes a practical set. Line of sight (LOS), fast steering mirror (FSM), optical telescope element (OTE), photoresponse nonuniformity (PRNU), clock-induced charge (CIC), charge-transfer inefficiency (CTI), and interpixel capacitance (IPC) denote distinct engineering interfaces. Surface height, reflected optical-path WFE, per-segment displacement, and modal RMS are not interchangeable: normal-incidence reflection doubles surface-height error. A segment-vector allocation requires its explicit basis and covariance, rather than identifying a scalar with every segment's piston.

\begin{table}[t]
\caption{Representative disturbance coordinates for HWO coronagraph error budgeting and typical units.
The Jacobians $J_k$ are provided by integrated modeling around the post-WFSC bias state.}
\label{tab:disturbance_coords}
\setlength{\tabcolsep}{5pt}
\begin{tabular}{p{0.22\linewidth} l p{0.54\linewidth}}
\toprule
Coordinate family & Units & Examples / notes \\
\midrule
Global pupil phase (Zernike) & nm RMS or rad & Low-order drift, thermal figure change; often controlled by HOWFS/HOWFSC \\
Segment rigid-body & pm--nm, nrad & Piston/tip/tilt per segment; includes differential segment motion and phasing residuals \\
Pointing / LOS & mas or $\lambda/D$ & FSM + LOWFS rejection; drives leakage through coronagraph sensitivity to tip/tilt \\
Pupil shear / magnification & \% of pupil & Lyot stop alignment, telescope pupil motion; can be treated as beamwalk state \\
DM drift states & nm surface & Creep/thermal drift of DM surfaces; couples through coronagraph response matrix \\
Polarization states & rad, fractional & Retardance/diattenuation modes mapped to effective phase/amplitude aberrations (Sec.~\ref{sec:pol}) \\
Detector / electronics & e$^{-}$, e$^{-}$/pix/frame, \% &
QE / PRNU, gain or threshold drift, CIC, dark current, CTI / radiation damage, nonlinearity, IPC; enters
$\bm\Sigma_{\rm det}$ and $\bm\Sigma_{\rm cal}$ rather than only the optical Jacobian branch \\

Optical bench / mechanism rigid-body & $\mu$m, $\mu$rad, pm &
Coronagraph optical-bench translation/clocking, focal-plane-mask and Lyot-stop decenter, precision-alignment
mechanism repeatability, thermoelastic bench modes; maps into pointing, pupil, and beamwalk coordinates \\

Spacecraft LOS / micro-vibration & mas, Hz-domain PSD &
Reaction-wheel tones, FSM residuals, line-of-sight drift, slew-settle transients, payload--spacecraft coupling;
shapes $S_{x_k}(f)$ and therefore $\Var(\Delta x_k)$ in Eq.~\eqref{eq:PSD_diff} \\
Stray light / ghosts & NI floor & Field or intensity terms according to mutual coherence; mean leakage and its differential change both enter \\
\bottomrule
\end{tabular}
\end{table}

\subsection{Scalar allocation: phase-blind bound and spatial overlap}
\label{sec:piston_example}

Retain the representative scalar sensitivity $S=6.90\times10^{-7}$ NI nm$^{-2}$, coherent level $\Ccoh=2\times10^{-10}$ NI, and suballocation $C_{\rm alloc}=3\times10^{-12}$ NI. These are declared allocation-test inputs, not a verified EAC1 segment-piston requirement. A spatially constant field $E_0=\sqrt{\Ccoh}$ and Jacobian $J=\sqrt S e^{i\theta}$ makes the missing phase dependence explicit. For centered Gaussian single-visit states of standard deviation $\sigma$ and correlation $\rho_v$, the exactly linear field gives
\begin{align}
\Var(\Delta I)&=8(1-\rho_v)\Ccoh S\cos^2\theta\,\sigma^2
+4(1-\rho_v^2)S^2\sigma^4\equiv a_2\sigma^2+a_4\sigma^4,\label{eq:scalar_exact}\\
\sigma_{\rm allow}^2&=\frac{2C_{\rm alloc}^2}{a_2+\sqrt{a_2^2+4a_4C_{\rm alloc}^2}}.
\label{eq:scalar_inverse}
\end{align}
The rationalized inverse is stable for a small quadratic term. For independent visits the conditional allowances are 0.09029, 0.12768, 0.51604, and 1.47442 pm at $\theta=0^\circ,45^\circ,80^\circ,90^\circ$; for $\rho_v=0.9$ they are 0.28548, 0.40357, 1.47809, and 2.23323 pm. In a blind search with uncontrolled phase, the relevant bound is the \emph{smallest} allowance. At fixed overlap, the phase-averaged linear surrogate is at most $\sqrt2$ optimistic in allowable RMS; the exact quadratic correction here is small. A quadrature relaxation requires demonstrated alignment and maintenance of $E_0$ relative to the relevant drift Jacobians.

\begin{figure}[t]
\centering
\includegraphics[width=.42\linewidth]{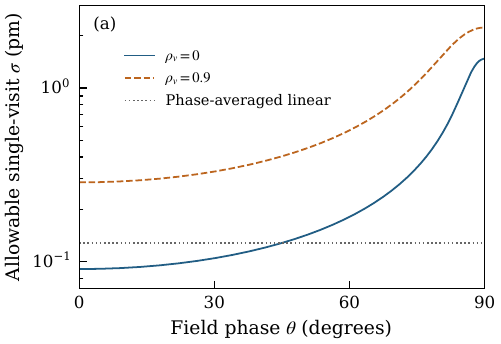}
\hskip 10pt 
\includegraphics[width=.42\linewidth]{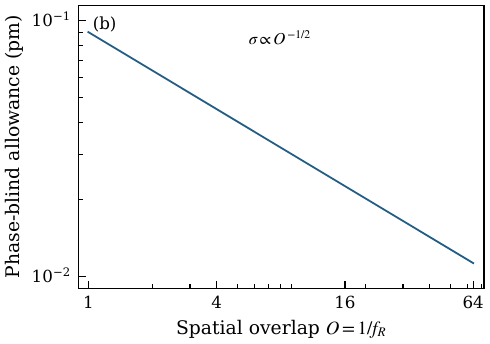}
\caption{Scalar allocation sensitivity for $\Ccoh=2\times10^{-10}$ NI, $S=6.90\times10^{-7}$ NI nm$^{-2}$, and $C_{\rm alloc}=3\times10^{-12}$ NI. (a) Phase-conditional allowances retaining quadratic intensity. The uncontrolled-phase requirement is the minimum, not the quadrature endpoint. (b) Aligned fields concentrated into a fraction $1/O$ of the region at fixed regional $\Ccoh$ and $S$. The approximately $O^{-1/2}$ tightening is accompanied by increasing local leakage; the example is a map-concentration test, not an HWO optical map.}
\label{fig:phase_allocation}
\end{figure}

Overlap can impose a larger penalty. For $N_R$ equal-area samples, $0\le O_k\le N_R$, while a sampling-invariant local bound is
\begin{equation}
O_k\le\min\left(\frac{\max_R|E_0|^2}{\Ccoh},\frac{\max_R|J_k|^2}{S_k}\right).
\label{eq:overlap_bound}
\end{equation}
Refining the sampling of a fixed map does not change $O_k$. To exercise this factor, let aligned fields $E_0=\sqrt{\Ccoh/f_R}$ and $J=\sqrt{S/f_R}$ occupy a common fraction $f_R$ of the region and vanish elsewhere. Then $O=1/f_R$, $P=1$, and the regional mean variance for independent visits is
\begin{equation}
\langle\Var(\Delta I)\rangle_R=8\Ccoh SO\sigma^2+4S^2O\sigma^4.\label{eq:overlap_example}
\end{equation}
For $O=1,4,16,64$, the same regional allocation permits 0.09029, 0.04514, 0.02257, and 0.01129 pm. The last case raises local bias intensity to $1.28\times10^{-8}$ NI and would fail a $10^{-10}$-class pointwise dark-hole constraint. Figure~\ref{fig:phase_allocation} separates the bounded phase effect from this concentration effect: both the regional allocation and local leakage constraint matter.

\subsection{Category-level allocation structure}\label{sec:categories}
The twelve contrast-stability source families identified by Liu et al.\ \cite{Liu2026EAC} organize this budget: full-aperture Zernike drift; changes in tip--tilt, segment, and WFE jitter; residual segment disturbances; low-order WFSC (LOWFSC) sensing, actuation, and camera stability; pointing repeatability; DM thermal drift and creep; pupil shear; and OTE and internal-coronagraph beamwalk. A thirteenth source-size/chromatic-calibration category is added here and retains its own physical source model.

As an arithmetic allocation test, assign thirteen zero-mean projected categories $5\times10^{-12}$ NI each. The resulting $\sqrt{13}\,5\times10^{-12}=1.8028\times10^{-11}$ NI is \emph{our illustrative allocation}, not a number taken from the project source table. With common pairwise correlation $r_c$,
\begin{equation}
C_{\rm total}=5\times10^{-12}\sqrt{13(1+12r_c)},\qquad -1/12\le r_c\le1.
\label{eq:category_correlation}
\end{equation}
A modest common correlation, $r_c=0.1$, raises the total to $2.674\times10^{-11}$ NI because every category pair contributes. Dividing the same physical disturbance among more names does not recover the independent RSS margin. The scalar $3\times10^{-12}$ suballocation represents a standard-deviation fraction $3/(5\sqrt{13})$ of the independent total; Sec.~\ref{sec:closed_case} preserves that fraction when the count budget is imposed. Before doing so, the residual amplitudes must be related to the time intervals on which calibration and control operate.

\section{Temporal domains, WFSC rejection, and PSD-based closure}
\label{sec:temporal}

An optical sensitivity specifies the instantaneous response to a disturbance; the observing window determines how that response contributes to the measurement. Rapid fluctuations can average within a visit, while slowly varying components can cancel between nearby visits. Intermediate timescales can therefore matter more than either extreme. To connect a control-system specification to contrast stability, we distinguish the instantaneous residual variance from the variance of the calibrated visit difference.

The instantaneous closed-loop variance of coordinate $x_k$ is
\begin{equation}
\sigma_k^2=\int_0^\infty |H_k(f)|^2\,S_{x_k}(f)\,df,
\label{eq:PSD}
\end{equation}
where $S_{x_k}$ is the one-sided environmental PSD and $H_k$ the residual control transfer function.

For stationary residuals observed in two visits of duration $T$ and midpoint spacing $\Delta t$, contrast stability instead uses the variance of their averaged difference, $\Delta x_k=\bar x_{k,A}-\bar x_{k,B}$:
\begin{equation}
\Var(\Delta x_k)
=
\int_{0}^{\infty} |H_k(f)|^2\,S_{x_k}(f)\;
\left|\mathrm{sinc}(\pi f T)\right|^2\;
4\sin^2(\pi f \Delta t)\,df,
\label{eq:PSD_diff}
\end{equation}
where the $\mathrm{sinc}^2$ (with $\mathrm{sinc}(u)\equiv \sin(u)/u$) factor represents within-visit averaging and the $4\sin^2(\pi f\Delta t)$ factor represents target--reference or roll differencing.

Eq.~\eqref{eq:PSD_diff} is therefore the observation-level temporal closure that should be used in
Eqs.~\eqref{eq:Cstab_single}--\eqref{eq:sigma_alloc} when converting disturbance environments into stability allocations. Very low-frequency drift ($f\ll 1/\Delta t$) is strongly suppressed by differencing, while mid-frequency content near $1/\Delta t$ is passed efficiently and often dominates the stability budget after WFSC filtering
\cite{Nemati2023RomanErrorBudget}.

\subsection{Multivariate temporal closure with cross spectra}
\label{sec:multivariate_temporal}

For a vector of residual disturbance coordinates $\bm x(t)$ driven by an environment/state vector
$\bm u(t)$, let the closed-loop residual transfer matrix be
\begin{equation}
\tilde{\bm x}(f)=\Hmat(f)\,\tilde{\bm u}(f),
\qquad
\bm S_x(f)=\Hmat(f)\,\bm S_u(f)\,\Hmat^\dagger(f),
\label{eq:Sx_matrix}
\end{equation}
where $\bm S_u(f)$ is the cross-spectral-density matrix of the driving processes and may be non-diagonal.

The covariance of the differenced, time-averaged residual state between two observations of duration $T$
separated by $\Delta t$ is then
\begin{equation}
\SigmaDx=\Ree\int_0^\infty W(f;T,\Delta t)\bm S_x(f)\,df,\qquad
W=|\sinc(\pi fT)|^2\,4\sin^2(\pi f\Delta t).
\label{eq:SigmaDx_matrix}
\end{equation}
Equation~\eqref{eq:PSD_diff} is recovered as the scalar, diagonal case. The real part is required for the real zero-lag covariance; complex cross spectra retain phase lags before integration. For unequal or interrupted visits replace $W$ by the product of the Fourier transforms of the actual normalized observation windows. Narrow vibration lines require their integrated line power, and deterministic slew settling belongs in a time-dependent mean rather than a stationary PSD.

Under coherent-mixing dominance, the corresponding intensity covariance is
\begin{equation}
\SigmaDI(\lambda)
=
\Amat(\lambda)\,\SigmaDx\,\Amat^{\mathsf T}(\lambda)
+
\SigmaInc(\lambda),
\label{eq:SigmaDI_full}
\end{equation}
Here $\bm A$ is defined by Eq.~\eqref{eq:A_matrix} and $\SigmaInc$ contains separately modeled intensity-only drifts satisfying the coherence accounting of Sec.~\ref{sec:raw_contrast}. The sum assumes zero cross covariance between branches; otherwise joint propagation is required. Shared polarization responses and coherent fluctuation power use Eqs.~\eqref{eq:gaussian_moments} and \eqref{eq:temporal_quadratic}, rather than being classified as independent intensity noise.

The regional stability follows from Eq.~\eqref{eq:Cstab_trace}, using this observation-weighted covariance and its differential mean.
The temporal model therefore supplies both the residual covariance and its observing-sequence dependence; the regional RMS is a summary of those products, not a substitute for them.

\subsection{Closed-loop residuals including sensor noise and discrete-time effects}\label{sec:closed_loop_residuals_including_sensor_noise_and_discrete_time_effects}

Eq.~\eqref{eq:PSD_diff} describes disturbance rejection but does not yet include sensing noise, delays, or sampling.
A more complete single-input single-output representation is
\begin{equation}
S_{x_k}^{\rm res}(f)\;=\;\left|H_{d,k}(f)\right|^2 S_{x_k}^{\rm env}(f)\;+\;\left|H_{n,k}(f)\right|^2 S_{n_k}(f),
\label{eq:residual_noise_spectrum}
\end{equation}
where $S_{x_k}^{\rm env}$ is the environment PSD, $S_{n_k}$ is the equivalent measurement (or estimation) noise PSD
mapped into the $x_k$ coordinate, and $H_{d,k},H_{n,k}$ are the disturbance- and noise-to-residual transfer functions.
The additive spectrum assumes independent environmental and sensing inputs. Shared drivers require the corresponding $H_dS_{un}H_n^\dagger+\mathrm{h.c.}$ terms. For a standard negative-feedback loop with open-loop gain $L_k(f)$,
\begin{equation}
H_{d,k}(f)=\frac{1}{1+L_k(f)},\qquad
H_{n,k}(f)=-\frac{L_k(f)}{1+L_k(f)}.
\label{eq:feedback_transfer}
\end{equation}
Increasing loop gain suppresses environmental motion while admitting more sensing noise. Both paths must enter the residual spectrum, as in Roman-derived error budgeting \cite{Nemati2023RomanErrorBudget,NematiStahl2024HWOErrorBudgetRoman}.

\subsection{Optical-assembly, rigid-body, and spacecraft-coupled state vector}
\label{sec:assembly_states}

Several disturbance coordinates in Table~\ref{tab:disturbance_coords} are not fundamental observatory states but
derived optical coordinates. A convenient intermediate representation is an assembly/spacecraft state vector
\begin{equation}
\bm q(t)=
\big[
\theta_x,\theta_y,\,
\delta x_{\rm pup},\delta y_{\rm pup},\delta\phi_{\rm pup},\,
\delta x_{\rm FPM},\delta y_{\rm FPM},\,
\delta x_{\rm LS},\delta y_{\rm LS},\,
q^{\rm bench}_1,\ldots,q^{\rm bench}_{N_b},\,
q^{\rm DM}_1,\ldots
\big]^{\mathsf T},
\label{eq:q_state}
\end{equation}
where the components may represent line-of-sight jitter/drift, pupil translation/clocking, focal-plane-mask and Lyot-stop
decenter, optical-bench thermoelastic modes, mechanism repeatability, and DM-package rigid-body or thermal states.

The coronagraph disturbance coordinates used by the Jacobians are then linearized as
\begin{equation}
\Delta\bm x \;=\; \bm B\,\Delta\bm q,
\label{eq:B_map}
\end{equation}
so that the covariance and PSD of the optical disturbance vector follow as
\begin{equation}
\bm\Sigma_{\Delta x} = \bm B\,\bm\Sigma_{\Delta q}\,\bm B^{\mathsf T},
\qquad
\bm S_x(f)=\bm B\,\bm S_q(f)\,\bm B^{\mathsf T}.
\label{eq:Sx_from_Sq}
\end{equation}

The map $\bm B$ connects mechanically specified states to optical sensitivity coordinates. Roman bench, alignment, and HOWFSC studies illustrate these interfaces \cite{Creager2025RomanBench,Krause2025RomanPAM,Cady2025RomanHOWFSC}; Tesch et al. describe HWO control development \cite{Tesch2026WFSC}.

\subsection{Mean and variance states for pointing, jitter, and upstream beamwalk}
\label{sec:mean_variance_states}

Two visits can have the same RMS pointing error but different image centroids or different jitter broadening. Those changes affect subtraction differently, and upstream beamwalk need not disappear when the focal-plane centroid is stabilized. For each axis $a\in\{x,y\}$, we therefore define separate visit-level mean and variance states
\begin{equation}
M_a \equiv \frac{1}{T}\int_0^T \theta_a(t)\,dt,
\qquad
V_a \equiv \frac{1}{T}\int_0^T \Big[\theta_a(t)-M_a\Big]^2 dt.
\label{eq:MV_defs}
\end{equation}
The differenced intensity may then be linearized as
\begin{equation}
\Delta \bm I
\simeq
\bm A_M\,\Delta\bm M
+
\bm A_V\,\Delta\bm V,
\qquad
[\bm A_M]_{qa}=\frac{\partial I_q}{\partial M_a},
\quad
[\bm A_V]_{qa}=\frac{\partial I_q}{\partial V_a},
\label{eq:MV_linear}
\end{equation}
so that
\begin{align}
\bm\Sigma_{\Delta I,{\rm jitter}}
&=
\bm A_M\,\bm\Sigma_{\Delta M}\,\bm A_M^{\mathsf T}
+
\bm A_V\,\bm\Sigma_{\Delta V}\,\bm A_V^{\mathsf T}
+
\bm A_M\,\bm\Sigma_{\Delta M,\Delta V}\,\bm A_V^{\mathsf T}
+
\bm A_V\,\bm\Sigma_{\Delta V,\Delta M}\,\bm A_M^{\mathsf T}.
\label{eq:Sigma_MV}
\end{align}
This form separates
(i) a mean-pointing repeatability requirement,
(ii) a change-in-jitter requirement between visits or rolls, and
(iii) beamwalk terms introduced upstream of the FSM that are not removed by focal-plane stabilization.

If one wishes to embed these states in the assembly vector of Eq.~\eqref{eq:q_state},
a convenient extension is
\begin{equation}
\bm q_{\rm ext}
=
\big[
\bm q^{\mathsf T},
M_x,M_y,V_x,V_y
\big]^{\mathsf T},
\label{eq:augmented_pointing_state}
\end{equation}
with the appropriate augmented mapping matrix replacing Eq.~\eqref{eq:B_map}.

\subsection{WFSC residual-state representation in mean and variance coordinates}
\label{sec:wfsc_mv_general}

Equations~\eqref{eq:MV_defs}--\eqref{eq:Sigma_MV} generalize to each controlled $x_k$ by replacing $\theta_a$. Use $\Delta M_k=M_{k,A}-M_{k,B}$ and $\Delta V_k=V_{k,A}-V_{k,B}$; conventional RMS is the derived $\sqrt{M_k^2+V_k}$. This realizes the project's four-statistic field description in disturbance coordinates \cite{Liu2026EAC}. Mean changes modify coherent mixing, while variance changes modify quadratic fluctuation power without creating a physically incoherent source.

The variance-state representation must retain cross-mode within-visit covariance when it is present. For normalized visit window $w_v(t)$, define $\bm M_v=\int w_v\bm x\,dt$ and $\bm V_v^{\rm within}=\int w_v(\bm x-\bm M_v)(\bm x-\bm M_v)^Tdt$. Equation~\eqref{eq:quadratic_intensity} then gives the pathwise identity
\begin{equation}
\overline I_{q,v}=c_q+\bm a_q^T\bm M_v+\bm M_v^T\bm Q_q\bm M_v+\Tr(\bm Q_q\bm V_v^{\rm within}).
\label{eq:visit_full_identity}
\end{equation}
Thus off-diagonal jitter/beamwalk variance and changes of $\bm M\bm M^T$ are not optional when their sensitivities are nonzero. The linearized mean/variance-state equations above are local derivatives of this identity, not a replacement for it. Augmented assembly and mean/variance coordinates describe the same physical motion and must not be allocated twice.
For a Gaussian process of mean $\bm\mu(t)$ and cross covariance $\bm K(t,t')$, direct averaging instead uses
\begin{equation}
K_{I,qr}(t,t')=\bm h_q(t)^T\bm K(t,t')\bm h_r(t')+
2\Tr[\bm Q_q\bm K(t,t')\bm Q_r\bm K(t',t)],
\label{eq:temporal_quadratic}
\end{equation}
where $\bm h_q(t)=\bm a_q+2\bm Q_q\bm\mu(t)$. Integrating this expression against the actual signed windows retains fluctuations in within-visit power without assuming independent sample means and variances.

\subsection{Example residual transfer functions and domain splits}
\label{sec:Hk_examples}

A useful screening model is a continuous integrator with unity-gain frequency $f_{c,k}$ and residual magnitude
\begin{equation}
|H_k(f)|=\frac{f}{\sqrt{f^2+f_{c,k}^2}},\qquad L_k(f)=\frac{f_{c,k}}{if}.
\label{eq:Hk_firstorder}
\end{equation}
This ideal continuous-time integrator rejects low-frequency drift while passing high-frequency jitter. Discrete implementations require sampling, aliasing, latency, actuator response, and stability margins; their residual spectra are integrated to the appropriate Nyquist frequency, with aliased input power included.

A practical error-budget implementation frequently partitions the disturbance PSD into temporal domains,
\begin{equation}
\sigma_k^2 \;=\; \sigma_{k,\rm drift}^2+\sigma_{k,\rm mid}^2+\sigma_{k,\rm jitter}^2,
\qquad
\sigma_{k,\rm drift}^2\equiv \int_0^{f_1}|H_k(f)|^2S_{x_k}(f)\,df,\ \ \text{etc.}
\label{eq:temporal_band_partition}
\end{equation}
where $(f_1,f_2)$ delimit bands relative to LOWFS/HOWFS bandwidths, roll spacing, and wavefront-control visits \cite{Nemati2023RomanErrorBudget,Liu2026EAC}. The allocation still uses the observation-weighted spectrum, not these bands' unweighted sum.

\begin{figure}[t]
\centering
\includegraphics[width=0.52\linewidth]{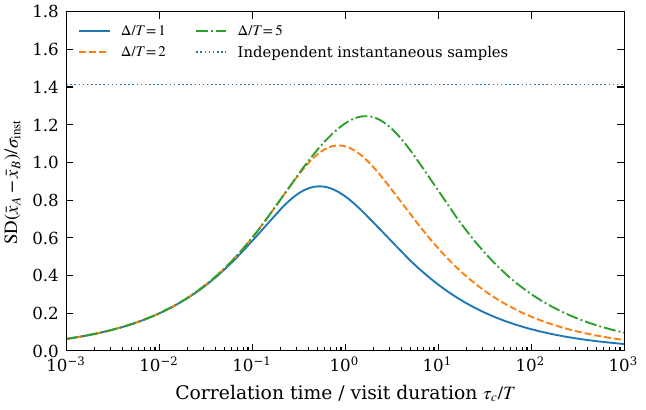}
\caption{Residual RMS of the difference between two finite-duration visits for the OU process in Eq.~\eqref{eq:ou_windows}, divided by instantaneous RMS. Curves hold the visit midpoint spacing at one, two, or five visit durations. The dotted $\sqrt2$ reference describes independent instantaneous samples, not these averages. The plotted process is already the closed-loop residual; no extra rejection factor is applied.}
\label{fig:temporal_windows}
\end{figure}

\paragraph*{Quantitative finite-window check.}
For an Ornstein--Uhlenbeck residual with $K_x(\tau)=\sigma_{\rm inst}^2e^{-|\tau|/\tau_c}$ and one-sided spectrum $4\sigma_{\rm inst}^2\tau_c/[1+(2\pi f\tau_c)^2]$, let $r=\tau_c/T$ and $d_v=\Delta t/T\ge1$. Direct integration of the same observing windows gives
\begin{align}
A(r)&=2[r-r^2(1-e^{-1/r})],\quad
B(r,d_v)=r^2e^{-(d_v-1)/r}(1-e^{-1/r})^2,\nonumber\\
\Var(\bar x_A-\bar x_B)&=\sigma_{\rm inst}^2D(r,d_v),\qquad D=2(A-B).
\label{eq:ou_windows}
\end{align}
For adjacent visits, the normalized differential RMS is 0.19849, 0.58310, 0.81998, 0.35183, and 0.11504 at $r=0.01,0.1,1,10,100$. The limiting forms $D(r,1)\simeq4r$ for $r\ll1$ and $D(r,1)\simeq4/(3r)$ for $r\gg1$ explain both sides of Fig.~\ref{fig:temporal_windows}: rapid fluctuations average within a visit, whereas very slow fluctuations are nearly common to both. Intermediate fluctuations do neither. Increasing visit separation weakens common-mode cancellation. This cancellation concerns a shared stationary random state; a deterministic change between visits remains in the differential mean.
For $I=c+\ell x+q x^2$ and centered OU $x$,
\begin{equation}
\Var(\bar I_A-\bar I_B)=\ell^2\sigma_{\rm inst}^2D(r,d_v)+2q^2\sigma_{\rm inst}^4D(r/2,d_v).
\label{eq:ou_quadratic_windows}
\end{equation}
For this centered Gaussian process, $\Cov[x^2(t),x^2(t+u)]=2\sigma_{\rm inst}^4e^{-2|u|/\tau_c}$. The centered squared fluctuation therefore has correlation time $\tau_c/2$. The mean of $x^2$ is positive, but its centered fluctuations still average, with a window response different from that of linear mixing. Assigning both terms the same rejection factor would mix these two effects. This distinction also carries into polarization, where mutually incoherent input channels can respond to the same time-dependent disturbance.

\section{Polarization-dependent aberrations}
\label{sec:pol}

\subsection{Unpolarized-input bookkeeping and polarization-channel closure}
\label{sec:unpolarized_bookkeeping}

Because stellar light at the entrance pupil is approximately unpolarized, the physically correct bookkeeping
is to propagate two mutually incoherent orthogonal polarization states through the optical train:
\begin{equation}
I_{\rm cam}(\bm\rho,\lambda)=\frac12\sum_{j=1}^2\|\bm E^{(j)}(\bm\rho,\lambda)\|^2,
\qquad \bm E^{(j)}=\Cop\{\bm J(\bm r,\lambda)\bm e_jP(\bm r)\}.
\label{eq:unpolarized_two_state}
\end{equation}
where $\bm e_1=(1,0)^{\mathsf T}$ and $\bm e_2=(0,1)^{\mathsf T}$ are any orthogonal Jones basis states.
The factor $1/2$ represents unit-total-flux input coherency $\bm I_2/2$; the norm sums detected orthogonal output components. Dividing the input fields by $\sqrt2$ as well would double-normalize. Observables are basis independent, but the two input channels remain mutually incoherent.

For small linear retardance $\eta(\bm r)$, diattenuation $d(\bm r)$, and fast-axis angle $\theta(\bm r)$,
the Jones matrix may be linearized as
\begin{equation}
\bm J(\bm r)\simeq\bm I+\frac{d(\bm r)+i\eta(\bm r)}2
\begin{bmatrix}\cos2\theta&\sin2\theta\\\sin2\theta&-\cos2\theta\end{bmatrix}.
\label{eq:J_small_pol}
\end{equation}
Here $d=(T_+-T_-)/(T_++T_-)$ is \emph{intensity} diattenuation, $\eta=\phi_+-\phi_-$ is the eigenphase difference, and common scalar transmission is already included in the reference response. The corresponding eigenchannel amplitude and phase scales are
\begin{equation}
a_{\rm pol,rms}\sim d_{\rm rms}/2,\qquad
\phi_{\rm pol,rms}\sim\eta_{\rm rms}/2,\qquad
\WFE_{\rm pol,rms}\sim\lambda\eta_{\rm rms}/(4\pi).
\label{eq:pol_amp_phase_scales}
\end{equation}
Pure lossless retardance produces eigenphase, not first-order eigenamplitude attenuation. In fixed axes it can nevertheless produce cross-polarized complex fields, so full vector propagation is required.

For the radial/tangential toy pattern $\eta(r)=\eta_p(r/R)^2$ and $\theta=\varphi$, take the unit-RMS real Zernikes $Z_6=\sqrt6(r/R)^2\cos2\varphi$ and $Z_5=\sqrt6(r/R)^2\sin2\varphi$. The fixed-axis co-polar phase coefficients and their difference are
\begin{equation}
a_{6,+}=\frac{\eta_p}{2\sqrt6},\quad a_{6,-}=-\frac{\eta_p}{2\sqrt6},\quad
a_{6,\rm pol}\equiv a_{6,+}-a_{6,-}=\frac{\eta_p}{\sqrt6}.
\label{eq:Z6_pol_mapping}
\end{equation}
The equivalent WFE coefficients are
\begin{equation}
|\WFE_{6,\pm}|=\frac{\lambda\eta_p}{4\pi\sqrt6},\qquad
\WFE_{6,\rm pol}=\frac{\lambda\eta_p}{2\pi\sqrt6}.
\label{eq:Z6_pol_WFE}
\end{equation}
At $\lambda=500$ nm and $\eta_p=10^{-4}$ rad these are 1.6244 pm per co-polar channel and 3.2487 pm differentially. The off-diagonal Jones term is simultaneously $i\eta_pZ_5/(2\sqrt6)$, so the full perturbation is not a single scalar astigmatism. The pattern's RMS eigen-retardance is $\eta_p/\sqrt3$, distinct from its peak.
This explicit mapping is useful when translating coating-derived peak retardance into Zernike coefficients
consumable by integrated coronagraph sensitivity models.

Polarization-dependent aberrations (PDA) therefore require both input coherency and the vector spatial response. An RMS \emph{eigen-retardance} $\eta_{\rm rms}=10^{-4}$ rad gives $\lambda\eta_{\rm rms}/(4\pi)=3.9789$ pm per eigenchannel at 500 nm, distinct from the peak-pattern Zernike coefficients above. Each channel's perturbation interferes with its bias; common disturbances can correlate the intensity residuals \cite{AshcraftSpencer2023PolAberr,Frazin2026Polarization}. The real unit-RMS Zernikes use the Noll convention \cite{Noll1976Zernike}.

\subsection{Linearized Jones model and leakage scaling in the dark hole}\label{sec:linearized_jones_model_and_leakage_scaling_in_the_dark_hole}
For a weak pupil-plane Jones perturbation $\bm J(\bm r)\simeq\bm I+\bm\epsilon(\bm r)$, $\|\bm\epsilon\|\ll1$, the matrix $\bm\epsilon$ carries diattenuation, retardance, and cross-polarization coupling. Its first-order field perturbation is
\begin{equation}
\Delta \bm E(\bm\rho,\lam)\;\approx\;\mathcal{C}\!\left\{\bm\epsilon(\bm r)\,\bm E_{\rm pup}(\bm r,\lam)\right\},
\label{eq:jones_field_response}
\end{equation}
with $\mathcal{C}\{\cdot\}$ denoting the coronagraph propagation operator.
For each independent input channel, this perturbation must be combined with that channel's bias field before intensities are summed. A contribution whose bias overlap vanishes can be quadratic; this is a property of the propagated fields, not a consequence of labeling it polarization leakage. Equation~\eqref{eq:gaussian_moments} then includes any shared disturbance covariance. Polarization-aware calculations establish the importance of carrying the vector response explicitly \cite{AshcraftSpencer2023PolAberr,Frazin2026Polarization}.

The response after control is an essential part of this accounting. Will et al.\ show limits of pairwise-type wavefront correction in the presence of polarization aberrations \cite{Will2025PolarizationLimits}; Ashcraft et al.\ connect vector physical-optics responses to HWO yield calculations \cite{Ashcraft2026PolarizationYield}. These results motivate carrying the attainable controlled field, polarization-dependent throughput, and common disturbance response together, rather than converting retardance into a universal independent contrast reserve.

\subsection{Diattenuation, cross-polarization leakage, and how to carry them in a stability budget}
\label{sec:pol_diattenuation}

In addition to retardance-driven polarization-dependent WFE, real coronagraph optical trains exhibit \emph{diattenuation} (polarization-
dependent throughput) and \emph{cross-polarization leakage} from coatings and fold mirrors. In a Jones description,
small departures from an ideal scalar pupil can be written as
\begin{equation}
\bm J(\bm r,\lam) \;=\;
\begin{bmatrix}
1+\epsilon_x(\bm r,\lam) & \gamma(\bm r,\lam) \\
\gamma'(\bm r,\lam) & 1+\epsilon_y(\bm r,\lam)
\end{bmatrix}
\exp\!\left[i\,\phi(\bm r,\lam)\right],
\label{eq:jones_perturbation}
\end{equation}
where $\epsilon_{x,y}$ represent diattenuation-like amplitude errors and $\gamma,\gamma'$ represent cross-pol terms.
To first order, $\epsilon$-terms behave like \emph{pupil amplitude aberrations} while $\phi$ is the usual phase
aberration; both can be budgeted through the same Jacobian machinery by mapping them into the appropriate $x_k$
coordinates used by the integrated model.

The engineering consequence is that coating quality and coating stability enter different parts of the measurement. A stable polarization-dependent halo raises the photon cost; a change between rolls can also create a planet-like residual. A retardance specification therefore needs the controlled vector response and the statistics of its temporal change, including correlation with scalar thermal WFE. With those quantities propagated in the same units as the other disturbances, the remaining step is to reconcile the optical budget with a particular target and observing program.

\section{From contrast stability to FRN and exposure time}
\label{sec:closure}

We now ask how much residual a specified observation can tolerate. A common target and count normalization distinguish an excessive optical allocation from an already photon-limited observation \cite{Nemati2020MethodPerformanceSpec,Nemati2023RomanErrorBudget}. We apply Eq.~\eqref{eq:FRN_speck} to a zero-mean signed-aperture residual, varying target and observing time at fixed response; Sec.~\ref{sec:signalextraction} supplies the general estimator treatment.

\subsection{A count-derived OS-1+ADI driving case}
\label{sec:closed_case}

Table~\ref{tab:closure_inputs} defines a clear 6-m \emph{photometric reference} and a Lambertian Earth analog at quadrature. The favorable 5-pc distance allows all retained bands through 1600 nm to pass the red-edge $3\lambda/D$ geometric screen. Instrumental responses are adopted inputs; target, total time, and extraction are held common across the comparisons.

\begin{table}[t]
\caption{Declared inputs for the count-derived closure and channel comparison. Instrument and background entries are adopted, not verified HWO performance. The top-hat bands, raw NI, and throughputs are listed in Table~\ref{tab:channel_alloc}; only the common reference efficiency is applied before those relative throughputs.}
\label{tab:closure_inputs}
\begin{tabular}{p{0.30\linewidth}p{0.54\linewidth}}
\toprule
Quantity & Value or definition \\
\midrule
Reference pupil; star distance & $D=6$ m; $A_{\rm col}=\pi D^2/4$; $d=5$ pc \\
Star & Blackbody, $R_\star=6.957\times10^8$ m, $T_\star=5772$ K \\
Planet & $R_p=6371$ km; $a=1$ au; $A_g=0.2$; $\alpha=90^\circ$ \\
Common reference response & $T_{\rm ref}\mathrm{QE}=0.20$; $\tau_s=1$ \\
Aperture & Angular radius $0.7\lambda_c/D$, fixed over each band \\
Total time; live fraction & $T_{\rm wall}=100$ h; $\eta_{\rm live}=0.80$ \\
Baseline extraction & Two equal rolls, both nonoverlapping lobes; no additional post-processing gain \\
Sky electron rate per aperture & $0.020(\Delta\lambda/100\,\mathrm{nm})(\lambda_c/500\,\mathrm{nm})^2$ s$^{-1}$ \\
Detector & Dark $0.001$ e$^-$ s$^{-1}$ per aperture; ideal counting, no read/CIC penalty \\
Calibration & Independent zero-mean 3.5-ppt program-level residual \\
Search model & $N_{\rm trial}=30000$; $P_{\rm FA,tot}=10^{-3}$; $f_{\rm miss}=0.01$; known Gaussian statistic \\
\bottomrule
\end{tabular}
\end{table}

The photon spectrum and normalization are
\begin{align}
F_\star(\lambda)&=\left(\frac{R_\star}{d}\right)^2\frac{2\pi c}{\lambda^4}
\left[\exp\!\left(\frac{hc}{\lambda k_BT_\star}\right)-1\right]^{-1},\label{eq:blackbody_rate}\\
\mathrm{PSF}_{\rm pk}(\lambda)&=\frac{\pi D^2}{4\lambda^2},\qquad
\bar g=\frac{\int R_\star(\lambda)\,[\pi^2(0.7)^2/4](\lambda_c/\lambda)^2d\lambda}{C_\star}.
\label{eq:band_geometry}
\end{align}
Equation~\eqref{eq:blackbody_rate} is per unit wavelength in meters; a per-nm integral includes its $10^{-9}$ Jacobian. Nominal solar constants follow Ref.~\cite{Prsa2016Constants}; a blackbody is an assumption, not a measured stellar spectrum. The sky-rate law fixes a uniform photon surface-brightness surrogate over each aperture and band, not an exozodiacal population. Detector performance, finite-star leakage, and continuous off-axis throughput must be replaced by the selected instrument's response before flight allocation.

For the 450--550 nm case, with $\tau_{\rm core}=0.12$ and $\Craw=3\times10^{-10}$ NI,
\begin{align}
C_\star&=2.3665863\times10^9\ \mathrm{e^-\,s^{-1}},&\quad \bar g&=1.2127344,\nonumber\\
C_p&=0.0327905\ \mathrm{e^-\,s^{-1}},& C_{\rm leak}&=0.8610122\ \mathrm{e^-\,s^{-1}},\nonumber\\
C_b&=0.0210\ \mathrm{e^-\,s^{-1}},&\kappac&=\bar g/\tau_{\rm core}=10.1061.
\label{eq:closed_rates}
\end{align}
The reference 20-ppt budget gives a mean SNR of 5.77, only slightly above the 5.40 search threshold. Statistical fluctuations therefore put many realizations below threshold: conditional power is 64.55\%. The specified 99\% power requires $\mathrm{SNR}_{\rm req}=7.72619$ and $\FRN_{\rm req}=14.9444$ ppt (Sec.~\ref{sec:false_alarm_and_margin}). This requirement follows from the chosen test, not from a universal optical constant. At 100 h the two-roll photon term is already 8.7953 ppt; including calibration leaves
\begin{equation}
\FRN_{\rm speck,allow}=\sqrt{14.9444^2-8.7953^2-3.5^2}=11.5641\ \mathrm{ppt}.
\label{eq:closed_allow}
\end{equation}
Equation~\eqref{eq:closed_allow} is the point at which the detection objective becomes an optical requirement: the allowance is the remainder after unavoidable photon noise and the declared calibration uncertainty, rather than a contrast value selected independently. For a residual uniform within the final planet-aligned extraction aperture, unit algorithm throughput, and no extra suppression, it maps to $C_{\rm stab,allow}=1.1443\times10^{-12}$ NI. Giving all of it to the $45^\circ$ scalar mode permits 48.70 fm; that is a whole-budget allowance, not a like-for-like comparison with a suballocation. Table~\ref{tab:allocation_shares} preserves the earlier amplitude fraction and separately shows an equal thirteen-way RSS share.
\begin{table}[t]
\caption{Hierarchy-preserving allocations within the same 5-pc, 100-h optical remainder. All rows use the stipulated scalar sensitivity, $O=1$, independent visits, and exact quadratic intensity. Fractions multiply standard deviation, not variance.}
\label{tab:allocation_shares}
\begin{tabular}{lrrrr}\toprule
Share & Fraction & $C_{\rm alloc}$ ($10^{-12}$ NI)&$\sigma_{45^\circ}$ (fm)&$\sigma_{0^\circ}$ (fm)\\\midrule
Entire optical remainder&1&1.1443&48.70&34.44\\
Earlier suballocation/total&$3/(5\sqrt{13})$&.1904&8.105&5.731\\
Equal thirteen-way RSS&$1/\sqrt{13}$&.3174&13.51&9.551\\\bottomrule
\end{tabular}
\end{table}

The illustrative $1.8028\times10^{-11}$-NI RSS would yield 182.19-ppt FRN \emph{if it described this same signed, uniform extracted residual}. Compatibility would then require $\eta_{\rm pp}/\eta_{\rm alg}\le0.06347$ or smaller allocations. This is a conditional consistency test, not a conclusion that the project budget fails. The EAC case is approximately 60 mas in a 20\% band at 600 nm and uses an off-axis PSF normalization \cite{Liu2026EAC}; this case is 200 mas at 500 nm in unocculted-peak NI. The driving tuple, response, and extraction must be reconciled first. Indeed, the adopted $3\lambda/D$ IWA is 61.88 mas at 600 nm and 68.07 mas at its red band edge: constant-throughput extrapolation to 60 mas would fail our own geometric screen.

Mean-repeatability comparisons require the projected mode response as well. The leading coherent response in Appendix~\ref{app:countcheck} is 159.28 ppt pm$^{-1}$. Retaining its quadratic correction, a 40-fm differential mean produces 6.3729-ppt bias. Multiplying the preliminary 0.2-pm integrated-model delta-mean goal reported by Liu et al.\ \cite{Liu2026EAC} by this coefficient gives 31.9 ppt before calibration. This identifies an interface to resolve: the stipulated $S$ has not been identified with that integrated-model coordinate or its estimator projection. Even a variance-only comparison to the entire 11.56-ppt optical remainder would require a response below 57.8 ppt pm$^{-1}$ at 0.2 pm; a signed mean instead consumes a separate bias allowance in the detection test.

\subsection{Target distance, luminosity, and observing-time boundaries}\label{sec:target_dependence}
Hold the visible response and the 5772-K spectral shape fixed while varying $\mathcal L=L_\star/L_\odot$, $d$, and wall time $T$. Set $R_\star=R_\odot\sqrt{\mathcal L}$ and an Earth-equivalent irradiation orbit $a=\sqrt{\mathcal L}$ au. This isolates luminosity in a constant-color family; actual stellar colors, finite-source leakage, and separation-dependent throughput require their own responses.

Let $z_d=(d/5\,\mathrm{pc})^2$, $\nu_0=C_{\star,0}\tau_{\rm core}$, and subscript zero denote the visible 5-pc benchmark. Then $f_p=f_{p,0}/\mathcal L$, $C_p=C_{p,0}/z_d$, and $C_{\rm leak}=\mathcal L C_{{\rm leak},0}/z_d$. The optical allowance is
\begin{equation}
C_{\rm stab,allow}=\frac1{\kappac}\left[\frac{\FRN_{{\rm req},0}^2}{\mathcal L^2}-s_{\rm cal}^2-
\frac{z_d(C_{p,0}+2\mathcal L C_{{\rm leak},0})+2C_bz_d^2}{\eta_{\rm live}T\nu_0^2\mathcal L^2}\right]^{1/2},\label{eq:target_allowance}
\end{equation}
for a positive radicand and an accessible planet. A negative radicand means infeasibility, not zero optical noise. With $a_d=C_{p,0}+2\mathcal L C_{{\rm leak},0}$ and $q_d=\eta_{\rm live}T\nu_0^2(\FRN_{{\rm req},0}^2-\mathcal L^2s_{\rm cal}^2)>0$,
\begin{equation}
z_{d,\max}=\frac{2q_d}{a_d+\sqrt{a_d^2+8C_bq_d}},\qquad d_{\rm rad}=5\sqrt{z_{d,\max}}\ \mathrm{pc}.\label{eq:distance_boundary}
\end{equation}
The independent red-edge geometric limit is $d_{\rm IWA}=17.6296\sqrt{\mathcal L}$ pc at 550 nm. Leakage-limited and sky-limited radiometric distances scale as $T^{1/2}$ and $T^{1/4}$, respectively. The calibration ceiling $\mathcal Ls_{\rm cal}\ge\FRN_{{\rm req},0}$ precludes any finite exposure.

\begin{figure}[t]
\centering\includegraphics[width=.40\linewidth]{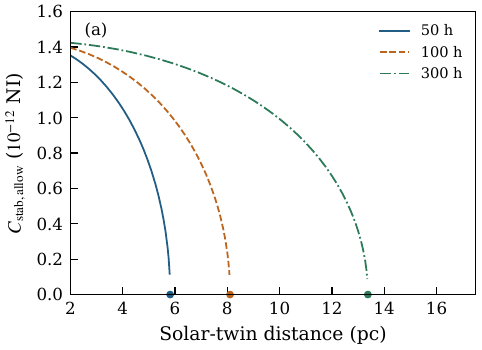}
\hskip 10pt
\includegraphics[width=.395\linewidth]{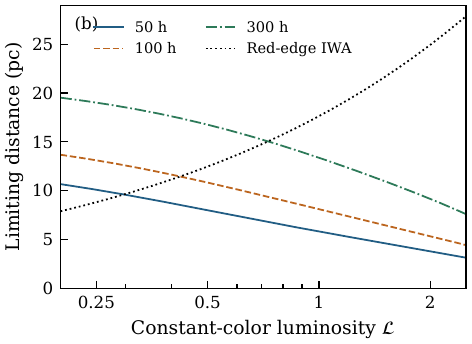}
\caption{Target dependence under the declared fixed-response model. (a) Solar-twin optical remainder at three total wall times; each curve ends where photon plus calibration noise exhausts the objective. (b) Radiometric distance boundaries versus constant-color luminosity, compared with the 550-nm $3\lambda/D$ geometric limit. Permitted distances are below both. The angular aperture, background rate, and calibration are fixed, while the orbit scales as $\sqrt{\mathcal L}$.}
\label{fig:target_dependence}
\end{figure}

For a solar twin the 100-h optical remainder vanishes at 8.1065 pc. At 10 pc the photon term alone is 18.197 ppt and the zero-optical-residual minimum time is 156.87 h; at 8 pc only $2.389\times10^{-13}$ NI remains. At 100 h, $(\mathcal L,d_{\rm rad},d_{\rm IWA})$ equals $(0.25,13.11,8.81)$, $(0.5,10.83,12.47)$, $(1,8.11,17.63)$, and $(2,5.36,24.93)$, with distances in pc. The two boundaries in Fig.~\ref{fig:target_dependence} correspond to different remedies. Inside the geometric limit but beyond the radiometric boundary, the planet is separated from the star, yet its available photons cannot meet the test at the allocated time even with zero optical residual. Better disturbance control alone cannot make that observation feasible. Conversely, a planet inside the IWA cannot be recovered merely by increasing exposure. In the constant-color family, lower luminosity improves flux ratio but moves the equal-irradiation orbit inward; the gain in radiometric reach therefore competes with a smaller angular-access limit. This is a scaling experiment, not a ranking of real stellar targets, whose colors and finite-source responses must also be included.

\subsection{Dominant terms, sensitivities, and equal-resource comparisons}
\label{sec:closure_sensitivity}

\begin{figure}[t]
\centering
\includegraphics[width=0.52\linewidth]{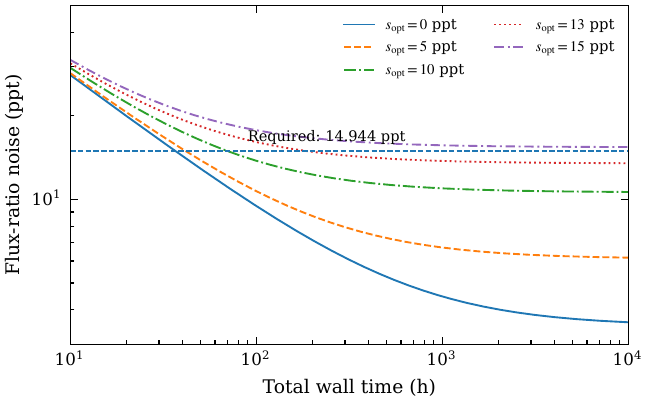}
\caption{OS-1+ADI flux-ratio uncertainty versus total wall time for the visible driving case in Table~\ref{tab:closure_inputs}. Each curve includes the same 3.5-ppt calibration term and a different persistent \emph{extracted} optical residual. The horizontal line is the 14.9444-ppt design objective. Reference/roll live time and the 80\% live fraction are included. The 15-ppt optical case cannot meet the objective even at infinite exposure. Curves are stationary-equivalent noise budgets, not an orbital observing forecast.}
\label{fig:closed_budget}
\end{figure}

For fixed throughput, sky, and detector parameters, the general closure is
\begin{equation}
\FRN^2(T)=\frac{V_{\rm ph}}{\eta_{\rm live}T(C_\star\tau_{\rm core})^2}+s^2,
\qquad s^2=\FRN_{\rm cal}^2+\FRN_{\rm speck}^2,
\label{eq:wall_closure}
\end{equation}
with $s<\FRN_{\rm req}$ necessary for finite time. Figure~\ref{fig:closed_budget} shows the separate allowances rather than assigning a universal contrast floor. With a 10-ppt optical residual and the 3.5-ppt calibration term, required total wall times are 35.45 h for an exactly known mean background, 69.64 h for OS-1+ADI, and 139.28 h for equal-time RDI. RDI is the hypothetical comparator defined in Sec.~\ref{sec:mode_indices}. These photon-cost cases hold the \emph{extracted} optical residual fixed; they do not establish that RDI and ADI attain that same residual in hardware. Real comparisons also recompute covariance, planet transmission, source mismatch, and overheads.

In the visible benchmark, stellar leakage accounts for 95.84\% of the photon variance coefficient. Thus reducing dark current alone has little leverage here, whereas throughput and mean stellar leakage matter even with excellent stability. At fixed reference normalization and uniform raw NI,
\begin{equation}
\frac{\partial\ln T}{\partial\ln C_{\rm raw}}=\frac{2C_{\rm leak}}{V_{\rm ph}},\quad
\frac{\partial\ln T}{\partial\ln\tau_{\rm core}}=-2+\frac{C_p}{V_{\rm ph}},\quad
\frac{\partial\ln T}{\partial\ln s}=\frac{2s^2}{\FRN_{\rm req}^2-s^2}.
\label{eq:closure_gradients}
\end{equation}
The first two derivatives are 0.9584 and $-1.9818$ at fixed extracted floor $s$. Stellar leakage dominates the numerator, while planet throughput enters the signal squared in the denominator. Thus a fractional reduction in leakage nearly gives the same fractional time saving; selectively increasing planet throughput has nearly twice the local leverage. This is not the scaling of a common transmission change, which also changes photon backgrounds (Sec.~\ref{sec:allocation_interpretation}). Figure~\ref{fig:closed_budget} shows a different loss mechanism: raising the persistent optical residual from 10 to 13 ppt, with calibration unchanged, increases ADI time from 69.64 to 183.81 h. At 14.5288 ppt no finite time suffices. The steeper cost is caused by exhaustion of the variance available for photon noise, rather than by extra photon arrivals.

The 100-h result is a stationary-equivalent calculation, not a multi-channel observing forecast. Real schedules propagate $f_p(t)$, projected separation, stellar variability, and moving backgrounds into the signal template. Table~\ref{tab:channel_alloc} recomputes band-integrated rates at the same time and distance rather than extrapolating the visible FRN. Per-channel SNR and molecular detectability are not interchangeable.

\section{Data analysis and signal extraction for imaging and spectroscopy}
\label{sec:signalextraction}

A different extraction can change both retained planet signal and residual uncertainty without changing the optical images. We therefore evaluate processing gains in inferred flux, not residual-image RMS, and distinguish imaging from spectroscopy objectives.

\subsection{A linear-operator view of calibration and post-processing}
\label{sec:linop}

Let $\bm d$ stack the calibrated images, time samples, or integral-field-spectrograph (IFS) data in a declared count or rate convention. We represent fixed calibration and subtraction operations by
\begin{equation}
\bm y = \bm M\,\bm d,
\label{eq:linop}
\end{equation}
where $\bm y$ is the processed data product used for detection or spectral extraction and $\bm M$ includes, as appropriate, ADI/roll-differenced subtraction (the baseline OS-1 branch), reference differential imaging (RDI) for alternate modes, spectral differencing, low-order background models, and/or a PCA/KL basis subtraction  \cite{Marois2006ADI,Soummer2012KLIP,Pueyo2016FMKLIP}. For a fixed linear operator, or a locally validated linearization, the covariance propagates as
\begin{equation}
\bm\Sigma_y = \bm M\,\bm\Sigma_d\,\bm M^{\mathsf T}.
\label{eq:covprop}
\end{equation}

Equation~\eqref{eq:covprop} shows why image suppression and improved inference need not coincide. A subtraction can remove a large common residual while attenuating the planet or correlating the remaining noise. The planet template and differential mean must therefore undergo the same transformation as the covariance. The scalar $\eta_{\rm pp}$ in Eq.~\eqref{eq:FRN_speck} is a summary of this operation, not an additional independent gain. If a principal-component-analysis or Karhunen--Lo\`eve (PCA/KL) basis is learned from the science data, its randomness and signal-dependent self-subtraction require forward modeling or injection/recovery of the complete procedure. Singular projected covariance is inverted only on its supported subspace.

\paragraph*{Baseline ADI and additional processing.}
The baseline uses $\bm M_{\rm base}=\bm M_{\rm ADI}$. Additional processing gives $\bm M=\bm M_{\rm COPP}\bm M_{\rm ADI}$, with $\bm M_{\rm COPP}\in\{\bm I,\bm M_{\rm spec},\bm M_{\rm alt}\}$ representing unchanged, spectral, or other processing. The transformed covariance and planet template determine the corresponding stability-relaxation ratio,
\begin{equation}
\mathcal R_{\rm COPP}(\lambda,\rho)=\frac{C_{\rm stab,allow}^{\rm (ADI+COPP)}(\lambda,\rho)}{C_{\rm stab,allow}^{\rm (ADI)}(\lambda,\rho)}.
\label{eq:Rcopp}
\end{equation}
with $\mathcal R_{\rm COPP} > 1$ indicating that the additional post-processing branch relaxes the observatory stability requirement at fixed science performance and total observing cost. A fixed linear transform cannot create information beyond the correctly modeled full-data GLS estimator; improvement over a specified nonoptimal baseline must not be reported as a gain over the optimal use of the same data.

\subsection{Detector, electronics, and calibration covariance terms}
\label{sec:detector_covariance}

Separate the unprocessed data covariance into its physical contributions:
\begin{equation}
\bm\Sigma_d
=
\bm\Sigma_{\rm ph}
+
\bm\Sigma_{\rm det}
+
\bm\Sigma_{\rm cal}
+
\bm\Sigma_{\rm speck},
\label{eq:Sigma_d_full}
\end{equation}
The terms describe photon statistics, stochastic detector effects, calibration uncertainty, and optical residuals in the same data space. This sum assumes zero cross covariance; shared drivers require joint propagation. The photon term uses unsubtracted mean counts.

For a linear extracted observable
\begin{equation}
y_i = \sum_q w_{iq}\,n_q,
\label{eq:linear_count_extraction}
\end{equation}
the leading calibration covariance terms, allowing cross-pixel correlations, are
\begin{align}
[\bm\Sigma_{\rm flat}]_{ij}&=\sum_{qr}w_{iq}w_{jr}\,\bar n_q\bar n_r\,\Cov(\delta q_q,\delta q_r),\label{eq:Sigma_flat}\\
[\bm\Sigma_{\rm gain}]_{ij}&=\sum_{qr}w_{iq}w_{jr}\,\bar n_q\bar n_r\,\Cov(\delta g_q,\delta g_r),\label{eq:Sigma_gain}\\
[\bm\Sigma_{\rm bg}]_{ij}&=\sum_{qr}w_{iq}w_{jr}\,\Cov(\delta b_{N,q},\delta b_{N,r}).\label{eq:Sigma_bg}
\end{align}
Here $\bar n_q$ is the mean in the \emph{same count units} as $n_q$, $\delta q_q$ and $\delta g_q$ are fractional response errors, and $\delta b_{N,q}$ is an additive error in counts (a rate error multiplied by its integration time). The diagonal-only formulas are special cases; cross correlations between flat, gain, background, and optical parameters must also be retained when present.

Nonlocal CTI, deferred charge, or interpixel coupling transforms the sample vector as $\bm n_{\rm obs}=\bm K_{\rm det}\bm n_{\rm ideal}$, so that
\begin{equation}
\bm\Sigma_{\rm obs}=\bm K_{\rm det}\bm\Sigma_{\rm ideal}\bm K_{\rm det}^T+\bm\Sigma_{\rm added}.
\label{eq:Sigma_nonlocal}
\end{equation}

Charge redistribution transforms existing fluctuations rather than necessarily adding independent noise: the difference $\bm K_{\rm det}\bm\Sigma_{\rm ideal}\bm K_{\rm det}^T-\bm\Sigma_{\rm ideal}$ need not be positive semidefinite. The transformed covariance, including any added detector noise, enters the same estimator as the optical residual \cite{Morrissey2023RomanEMCCD,Bush2025RomanCameraSystems}.

\subsection{Optimal detection and photometric estimators}
\label{sec:matchedfilter}

For a processed unit-flux planet template $\bm p$ in the data space of $\bm y$, the unknown flux amplitude $a$ is described by
\begin{equation}
\bm y = a\,\bm p + \bm n, \qquad \bm n\sim\mathcal{N}(\bm 0,\bm\Sigma_y).
\label{eq:gls_model}
\end{equation}
The generalized least-squares (GLS) / matched-filter estimator is
\begin{equation}
\widehat{a} = \frac{\bm p^{\mathsf T}\bm\Sigma_y^{-1}\bm y}{\bm p^{\mathsf T}\bm\Sigma_y^{-1}\bm p},
\qquad
\Var(\widehat{a}) = \left(\bm p^{\mathsf T}\bm\Sigma_y^{-1}\bm p\right)^{-1}.
\label{eq:gls}
\end{equation}
The corresponding detection significance (for a true amplitude $a$) is
\begin{equation}
\mathrm{SNR}_a = \frac{a}{\sqrt{\Var(\widehat{a})}} = a\,\sqrt{\bm p^{\mathsf T}\bm\Sigma_y^{-1}\bm p}.
\label{eq:snr_matched}
\end{equation}
The weighting in Eq.~\eqref{eq:gls} favors data combinations that contain planet signal with relatively little noise \cite{Ruffio2017MatchedFilter}. A bright residual mode that is distinct from the planet can be downweighted; a weaker mode aligned with the planet is more difficult to distinguish. This is the physical reason that equal per-pixel errors, or equal regional contrast RMS, need not imply equal flux uncertainty.

\paragraph*{Connection to FRN.}
The estimator's sampling standard deviation is the FRN defined in Sec.~\ref{sec:frn}:
\begin{equation}
\FRN(\lam) \equiv \sqrt{\Var\!\left(\widehat{f_p}(\lam)\right)}.
\label{eq:frn_estimator_definition}
\end{equation}
This recovers Eqs.~\eqref{eq:SNR_correlated}--\eqref{eq:t_req} for persistent aperture residuals; PSF fitting and multi-wavelength extraction use the full projected covariance with the same template normalization.

\paragraph*{Nuisance degeneracy, uncertainty, and model mismatch.}
Let the local data model be $\bm y=a\bm p+\bm B\bm\beta+\bm n$, with calibration/background templates $\bm B$ and independent Gaussian prior $\bm\beta\sim\mathcal N(0,\bm\Lambda)$. Marginalizing the calibration uncertainty gives $\bm C=\bm\Sigma_y+\bm B\bm\Lambda\bm B^T$ and
\begin{equation}
\bm W=\bm\Sigma_y^{-1}-\bm\Sigma_y^{-1}\bm B
(\bm B^T\bm\Sigma_y^{-1}\bm B+\bm\Lambda^{-1})^{-1}
\bm B^T\bm\Sigma_y^{-1}=\bm C^{-1}.
\label{eq:nuisance_precision}
\end{equation}
Using $\bm W$ in Eq.~\eqref{eq:gls} accounts for the ambiguity between the planet and the calibration templates. In the diffuse-prior limit, a nuisance pattern indistinguishable from the planet makes its amplitude unidentifiable; collecting more of the same data does not resolve that degeneracy. A finite prior supplies external calibration information, so its marginal variance averages over that declared calibration distribution rather than certifying every fixed calibration error.
For assumed-model weights $\bm w=\bm W\bm p/(\bm p^T\bm W\bm p)$, the actual variance and bias are
\begin{equation}
\Var_{\rm true}(\widehat a)=\bm w^T\bm\Sigma_{\rm true}\bm w,\quad
b_a=a(\bm w^T\bm p_{\rm true}-1)+\bm w^T\bm b_{\rm true},\quad
\mathrm{RMSE}^2=\Var_{\rm true}(\widehat a)+b_a^2.
\label{eq:estimator_mismatch}
\end{equation}
Estimated covariance is an additional uncertainty, not an exactly known inverse: covariance-parameter likelihoods retain the $\log\det\bm\Sigma$ term and should be checked by predictive coverage. Low counts require a Poisson/detector likelihood; nonlinear, boundary-limited molecular models need more than a local Gaussian error bar. Appendix~\ref{app:countcheck} provides a controlled example in which correct marginal sample variances but omitted spatial cross covariance produce undercoverage.

\subsection{Detection thresholds, false alarms, and why ``SNR requirement'' exceeds the threshold}
\label{sec:false_alarm_and_margin}

An uncertainty estimate answers how widely repeated flux estimates scatter. A detection test asks a different question: how often does that scatter cross a threshold when no planet is present, and how often does a real planet exceed it? These are the false-alarm probability and conditional detection power. Setting a design SNR equal to the threshold would place the mean planet signal at the decision boundary, giving only 50\% power for the symmetric Gaussian model. A stability requirement must therefore include a detection margin as well as a noise estimate.

\paragraph*{Single-trial and mission-level false-alarm probability.}

Let $z$ be the normalized detection statistic (e.g., the matched-filter amplitude normalized by its
standard deviation). Under the null hypothesis (no planet), $z$ is often modeled as standard normal:
$z\sim\mathcal{N}(0,1)$ after whitening.
For a threshold $z_{\rm th}$, the \emph{single-trial} false-alarm probability is
\begin{equation}
P_{\rm FA,1}(z_{\rm th}) \;=\; \frac{1}{2}\,\mathrm{erfc}\!\left(\frac{z_{\rm th}}{\sqrt{2}}\right).
\label{eq:PFA1}
\end{equation}
If the search involves $N_{\rm trial}$ approximately independent trials (distinct searched locations/templates with independently calibrated null statistics), then the \emph{family-wise} false-alarm
probability is
\begin{equation}
P_{\rm FA,tot} \;=\; 1-\left(1-P_{\rm FA,1}\right)^{N_{\rm trial}}
\;\simeq\; N_{\rm trial}\,P_{\rm FA,1}\quad(P_{\rm FA,1}\ll 1).
\label{eq:PFA_tot}
\end{equation}

For a filled annular dark hole spanning $\rho\in[\IWA,\OWA]$, $N_{\rm DH}\simeq\pi(\OWA^2-\IWA^2)$ is a geometric resolution-element estimate, not an independence proof. Define the trials factor by the hypotheses actually searched,
\begin{equation}
N_{\rm trial}=N_{\rm distinct\ search\ hypotheses}.
\label{eq:Ntrial_est}
\end{equation}
Wavelengths, rolls, and visits jointly fitted to one planet do not each create additional hypotheses. Arbitrary dependence can instead be controlled conservatively with $P_{\rm FA,1}=P_{\rm FA,tot}/N_{\rm trial}$ by the union bound, provided each marginal null is calibrated. Small samples, learned covariance, and selected templates can change the null distribution \cite{Mawet2014SmallSampleStats}. Whitening alone does not establish Gaussian tails.

\begin{figure}[t]
\centering\includegraphics[width=.40\linewidth]{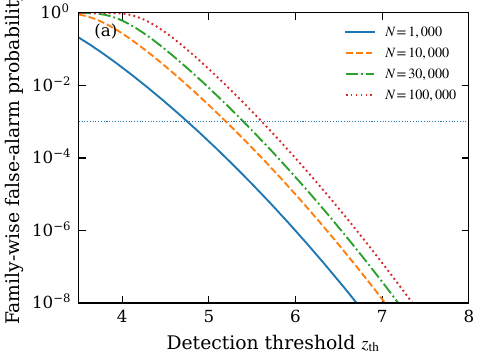}
\hskip 10pt
\includegraphics[width=.42\linewidth]{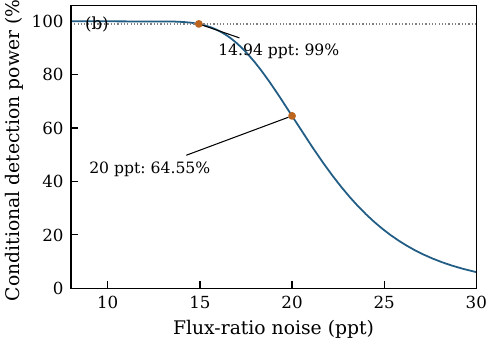}
\caption{Detection statistics used before allocation. (a) Family-wise false-alarm probability for independent Gaussian hypotheses. (b) Conditional detection power versus total FRN for the 115.4634-ppt planet, $N_{\rm trial}=30000$, and $P_{\rm FA,tot}=10^{-3}$. The same test needs 14.9444 ppt for 99\% power; 20 ppt gives 64.55\%. These are conditional test probabilities, not orbital or survey completeness.}
\label{fig:detection_statistics}
\end{figure}

Figure~\ref{fig:detection_statistics} links the family-wise threshold to conditional power at the resulting FRN.

\paragraph*{Missed-detection fraction and the SNR margin.}

Under the planet-present hypothesis with true matched-filter SNR $\mu\equiv \mathrm{SNR}_{\rm true}$, the normalized
statistic is approximately $z\sim\mathcal{N}(\mu,1)$.
The probability that an observation fails to meet the detection threshold is
\begin{equation}
f_{\rm miss} \;\equiv\; P(z<z_{\rm th}\,|\,\mu) \;=\; \Phi_{\rm N}(z_{\rm th}-\mu),
\label{eq:fmiss}
\end{equation}
where $\Phi_{\rm N}$ denotes the standard-normal cumulative distribution function (CDF), distinct from the Lambert phase function.
Solving for the required mean SNR gives
\begin{equation}
\mathrm{SNR}_{\rm req} \;=\; z_{\rm th} + \Phi_{\rm N}^{-1}(1-f_{\rm miss}).
\label{eq:SNRreq_margin}
\end{equation}
Thus, even if $z_{\rm th}$ is fixed by false-alarm control, the \emph{instrument} is typically designed for a larger
$\mathrm{SNR}_{\rm req}$ to ensure the stated conditional detection power.

\paragraph*{Connection to FRN allocations.}
The FRN requirement used for budgeting should therefore be interpreted as
\begin{equation}
\FRN_{\rm req}(\lam) \;\equiv\; \frac{f_p(\lam)}{\mathrm{SNR}_{\rm req}},
\label{eq:FRNreq_with_margin}
\end{equation}
not merely $f_p/\mathrm{SNR}_{\rm det}$, because $\mathrm{SNR}_{\rm req}$ encodes the desired missed-detection probability for that flux/geometry/noise model. Orbital and survey completeness further average over the planet population, orientation, accessible phases, and schedule.

For the numerical driving case, exact inversion of Eq.~\eqref{eq:PFA_tot} at $N_{\rm trial}=30000$ and $P_{\rm FA,tot}=10^{-3}$ gives $z_{\rm th}=5.39984$. Adding $\Phi_{\rm N}^{-1}(0.99)=2.32635$ gives $\mathrm{SNR}_{\rm req}=7.72619$ and $\FRN_{\rm req}=14.9444$ ppt for the 115.4634-ppt planet. In contrast, an assumed 20-ppt FRN gives only 64.55\% conditional power under this same test. The difference is not a new optical effect: it is the statistical cost of reliably recovering the planet rather than merely placing its mean signal above threshold.
If null and planet-present standard deviations differ, the exact Gaussian power condition is $f_p\ge z_{\rm th}\sigma_0+\Phi_{\rm N}^{-1}(1-f_{\rm miss})\sigma_1$. The numerical budget uses the larger design-planet variance for both, a conservative null normalization. Unknown signed bias shifts this threshold and must not be absorbed into FRN. A million ordinary realizations cannot verify single-trial false-alarm probabilities near $10^{-8}$; those require a justified tail model, rare-event methods, or independent empirical calibration.

\subsection{Constructing spatio-spectral covariance from Jacobians and disturbance statistics}
\label{sec:covariance_model}
Stacking wavelength with spatial sample in Eq.~\eqref{eq:gaussian_moments}, then applying the count and extraction operators to both residuals and planet, constructs the spatio-spectral covariance including quadratic terms.

For the linear-intensity limit, let $w_q$ be fixed extraction weights and $\nu_k(\lambda)=\sum_qw_q A_{qk}(\lambda)$, where $A_{qk}$ is defined in Eq.~\eqref{eq:A_matrix}. The extracted NI residual and its covariance are
\begin{align}
\Delta y(\lambda)&=\sum_k\nu_k(\lambda)\Delta x_k,
\label{eq:y_linear}\\
\Sigma^{\rm speck}_{ij}&=\sum_{k\ell}\nu_k(\lambda_i)\nu_\ell(\lambda_j)[\SigmaDx]_{k\ell}.
\label{eq:cov_speck_general}
\end{align}
The coefficients $\nu_k$ have units NI per unit of $x_k$; Sec.~\ref{sec:temporal} supplies $\SigmaDx$ from the residual cross spectra and observing windows. Independent coordinates are the diagonal special case, not a prerequisite for the propagation.

When the NI-to-rate operator is separable into a scalar $A(\lambda)$ after this extraction, the optical covariance in flux units is
\begin{equation}
\bm\Sigma_{\FRN}=\bm D\bm\Sigma^{\rm speck}\bm D,
\qquad\bm D=\diag\!\big(\kappa_{\FRN}(\lambda_i)A(\lambda_i)\big).
\label{eq:cov_FRN}
\end{equation}
Otherwise include the wavelength-dependent pixel areas and response inside the extraction. Add photon, detector, and calibration covariance in the same final data space, retaining their cross terms when present. The differential mean follows the same linear maps and supplies estimator bias separately.

\paragraph*{Imaging-mode extraction and photometry.}
\label{sec:imaging_extraction}
The imaging template $\bm p=\bm M\bm p_0$ includes planet self-subtraction calibrated by forward modeling or injection/recovery \cite{Soummer2012KLIP,Pueyo2016FMKLIP}. Its covariance and detection threshold refer to that same pipeline.

\subsection{Spectroscopy interface and channel-level accounting}
\label{sec:parametric_spectral_cov}
An imaging detection objective must not be imposed independently on every spectral element. For each bin, integrate the same reference and planet spectra:
\begin{equation}
C_{\star,b}=\int_bR_\star(\lambda)d\lambda,\qquad C_{p,b}=\int_bR_\star(\lambda)\tau_{\rm core}(\lambda)f_p(\lambda)d\lambda.\label{eq:CstarbCpb}
\end{equation}
Table~\ref{tab:channel_alloc} compares photon-plus-calibration precision at the same 100-h cost. Narrow bins receive fewer photons; even the 1600-nm broadband case is photon limited despite geometric access. Molecular information is distributed across a structured template, so single-bin continuum SNR alone does not determine feature detectability after joint extraction and nuisance marginalization.

For $\bm s=\bm M_s\bm a+\bm n$, let $\bm M_s$ contain instrument-convolved continuum and molecular derivative templates. The local Gaussian information is
\begin{equation}
\bm F=\bm M_s^T\bm\Sigma_s^{-1}\bm M_s,\qquad\Cov(\widehat{\bm a})=\bm F^{-1}.\label{eq:fisher}
\end{equation}
Nuisance parameters require the block inverse or Schur complement; covariance-dependent parameters additionally contribute $\tfrac12\Tr(\Sigma^{-1}\Sigma_{,i}\Sigma^{-1}\Sigma_{,j})$. Molecular boundaries and template searches require separately calibrated hypothesis tests.

The physical $\bm\Sigma_s$ follows from Eqs.~\eqref{eq:gaussian_moments} and \eqref{eq:cov_FRN} with the line-spread function and extraction applied once. A fitted kernel is a compression with calibrated amplitude, correlation length, and uncertainties. Chromatic magnification suggests $\Delta\lambda\sim\lambda/\rho$ only as an order-of-magnitude scale. High-pass filtering removes both broad residual modes and broad molecular information; it cannot create information relative to optimal use of the same data. Detailed spectral covariance and resolving-power analyses are given by Greco and Brandt and Ruffio et al. \cite{GrecoBrandt2016SpectralCovariance,Ruffio2025MHRS}. Here the deliverable is the physical covariance interface for those downstream analyses.
\begin{table*}[t]
\caption{Channel accounting for the 5-pc planet and 100-h total wall time. Band, throughput, and raw NI are adopted inputs. $\FRN_{\rm ph+cal}$ combines the calculated photon term and the independent 3.5-ppt calibration residual, with zero optical residual. The last column is continuum $f_p/\FRN_{\rm ph+cal}$, not molecular significance.}
\label{tab:channel_alloc}
\begin{tabular}{lrrrrrrr}\toprule
Channel&$\lambda_c$ (nm)&$\Delta\lambda/\lambda_c$&$\tau_{\rm core}$&$C_{\rm raw}$ ($10^{-10}$ NI)&$\FRN_{\rm rand}$ (ppt)&$\FRN_{\rm ph+cal}$ (ppt)&SNR\\\midrule
Visible broadband&500&0.2&0.12&3&8.80&9.47&12.20\\
NIR-1 broadband&1000&0.2&0.10&5&10.85&11.40&10.13\\
NIR-2 broadband&1600&0.2&0.07&8&24.71&24.95&4.63\\
Visible narrow band&800&0.007&0.07&3&63.82&63.92&1.81\\
NIR-1 narrow band&1000&0.007&0.06&5&96.25&96.31&1.20\\
NIR-2 narrow band&1600&0.014&0.05&8&130.33&130.38&0.89\\
\bottomrule\end{tabular}\end{table*}

\subsection{Scientific interpretation and model deliverables}
\label{sec:science_mapping}

Imaging and molecular estimators favor different disturbance reductions: the former rejects planet-like spatial residuals, while the latter rejects residuals aligned with spectral features after nuisance marginalization [Eq.~\eqref{eq:fisher}]. Table~\ref{tab:products_schema} supplies the response, mean, covariance, and unit interfaces for exposure-time and yield tools \cite{Stark2019YieldLandscape,Savransky2016EXOSIMS}. These products must be paired with target geometry, observing schedule, unsubtracted leakage, algorithm throughput, and calibrated search hypotheses. Alternative observing modes and compressed covariance kernels need their own response and validation. The following tests determine whether these reductions preserve the intended measurement.

\begin{table*}[t]
\caption{Minimum machine-readable products required for estimator-aware HWO coronagraph budget closure. Coordinate units and the input/output data spaces must accompany each array; NI maps and count-space products are not interchangeable.}
\label{tab:products_schema}
\setlength{\tabcolsep}{3pt}
\renewcommand{\arraystretch}{1.00}
\begin{tabular}{p{0.14\textwidth} p{0.16\textwidth} p{0.18\textwidth} p{0.40\textwidth}}
\toprule
Product & Dimensions & Units & Comment \\
\midrule
$\E_0^{(\Mmode,\Dmode)}(\lambda_i,\rho_q)$ &
$N_\lambda\times N_q$ &
$\sqrt{\mathrm{NI}}$ &
Coherent bias field after WFSC \\

$J_k^{(\Mmode,\Dmode)}(\lambda_i,\rho_q)$ &
$N_k\times N_\lambda\times N_q$ &
$\sqrt{\mathrm{NI}}/\mathrm{unit}(x_k)$ &
First-order field Jacobians \\

$\Gamma_{k\ell}^{(\Mmode,\Dmode)}(\lambda_i,\rho_q)$ &
selected pairs &
$\frac{\sqrt{\mathrm{NI}}}{\mathrm{unit}(x_k)\mathrm{unit}(x_\ell)}$ &
Second-order field sensitivities for linearity validation \\

$\tau_{\rm core}^{(\Mmode,\Dmode)}(\lambda_i,\rho)$ &
$N_\lambda\times N_\rho$ &
dimensionless &
Planet core throughput \\

$\kappac^{(\Mmode,\Dmode)}(\lambda_i,\rho)$ &
$N_\lambda\times N_\rho$ &
dimensionless &
NI-to-FRN mapping for the adopted aperture / PSF normalization \\

$\bm S_u(f_n)$ or $\SigmaDx$ &
state-space arrays &
state-units$^2\,\mathrm{Hz}^{-1}$ or state-units$^2$ &
Driving cross spectra or visit-level disturbance covariance \\

$\bm\mu_{\Delta I}$, $\bm\Sigma_{\Delta I}$ &
$N_s$; $N_s\times N_s$ &
NI; NI$^2$ &
Differential optical mean and covariance, $N_s=N_\lambda N_q$ \\

$\bm p$, $\bm M$ &
$N_y$; $N_y\times N_d$ &
data/flux ratio; operator &
Processed unit-flux template and declared data transformation \\

$\bm\Sigma_{\rm det}$, $\bm\Sigma_{\rm cal}$ &
estimator space &
flux-ratio$^2$ &
Detector and calibration covariance terms \\

$\bm\Sigma^{(\Mmode,\Dmode)}(\lambda_i,\lambda_j;\rho)$ &
$N_\lambda\times N_\lambda\times N_\rho$ &
flux-ratio$^2$ &
Final covariance consumed by ETC / yield / retrieval tools \\
\bottomrule
\end{tabular}
\end{table*}

\section{Implementation and validation}
\label{sec:validation}
Agreement within an assumed model verifies implementation, not every physical assumption. We separately test optical truncation and disturbance distributions at fixed target, estimator, and observing cost, then change the available state information explicitly. Harmonizing these inputs also makes cross-model differences interpretable \cite{Stark2025ETCValidation,Steiger2026EBS}.

\subsection{Normalization verification and nonlinear optical completion}
\label{sec:completed_checks}
Independent quadratures check photon rates and OU variances; Zernike projection checks polarization coefficients, and aperture subdivision checks count invariance. Appendices~\ref{app:countcheck} and \ref{app:numerical} specify the Gaussian field/count self-test, covariance-sign reversal, and draw order. These verify implementation rather than flight optical accuracy.

To test the linear-field assumption at the allocation scale, complete the local field as a two-path response,
\begin{equation}
E_q^{\rm nl}(x)=E_{0,q}+\frac{J_q}{ik}(e^{ikx}-1),\qquad k=\frac{2\pi}{500\ \mathrm{nm}}.
\label{eq:nonlinear_completion}
\end{equation}
This has the same $E_0$ and $J$ as the appendix model but a definite nonlinear phase dependence. Writing it as $a_q+b_qe^{ikx}$ gives $\max_q(|a_q|+|b_q|)<0.168$, compatible with a passive attenuated two-path realization. At the 8.105-, 13.51-, and 48.70-fm scales of Table~\ref{tab:allocation_shares}, and at 80 fm, the linear-field optical variance differs from this nonlinear propagation by at most $1.35\times10^{-7}$ fractionally. The maximum bias difference over these four scales is $4.00\times10^{-7}$ ppt. Sixteen- and thirty-two-node Gaussian product quadrature agree to better than $10^{-11}$ in relative optical variance. The small discrepancy is expected: at 80 fm, $k\sigma=1.01\times10^{-6}$ rad, so the first omitted field term is small relative to the nonzero Jacobian. This is a local consistency check, not a demanding test of response order.

A symmetry-protected null tests a different issue. For an ideal clear circular pupil, on-axis point source, charge-6 vortex, and circular Lyot stop, a pupil term $Z_n^m$ is rejected when $n+|m|<6$ \cite{Ruane2017VortexSegmented}. In the expansion of a pure tip--tilt phase $\exp[iu(r/R)\cos\phi]$, the first transmitted term is cubic, containing $n=|m|=3$. Thus $E=\mathcal O(u^3)$ and leakage is proportional to $u^6$: both $J$ and $\Gamma$ vanish at the origin, so even a squared quadratic field misses the leading response. Pointing bias, finite stellar size, segmentation, and other symmetry-breaking errors alter that order. Each actual mode requires validation of its first nonzero response, not merely a small field-norm error.

\subsection{Intermittency at fixed disturbance mean and covariance}
\label{sec:distribution_stress}
A small optical truncation error does not guarantee a correct detection threshold. The latter also depends on how often the observatory occupies states far from its usual residual amplitude. To isolate that issue, we retain the two-aperture response, 100-h count budget, $\bm\mu=(0.025,-0.015)$ pm, inter-visit correlation 0.6, and independent 3.5-ppt calibration residual, but replace the Gaussian state by
\begin{equation}
\bm x=\bm\mu+\sqrt W\,\bm L\bm u,\quad\bm u\sim\mathcal N(0,\bm I),\quad\bm L\bm L^T=\bm V,
\qquad W=\begin{cases}(1-f_v)/(1-p_e),&\text{probability }1-p_e,\\f_v/p_e,&\text{probability }p_e.\end{cases}
\label{eq:mixture_state}
\end{equation}
Here $p_e$ is the probability of a high-variance observing program and $f_v$ its share of the ensemble disturbance variance. These are stress parameters, not measured HWO event rates; the model describes intermittency of visit-averaged states, not an intra-visit waveform. Since $\Eavg W=1$, the disturbance mean and covariance are unchanged. Its fourth moments change through $\Eavg W^2=(1-f_v)^2/(1-p_e)+f_v^2/p_e$. For $y=\bm a^T\bm x+\bm x^T\bm Q\bm x$,
\begin{equation}
\Var y=\bm h^T\bm V\bm h+2\Eavg(W^2)\Tr(\bm Q\bm V\bm Q\bm V)
+\Var(W)[\Tr(\bm Q\bm V)]^2,\quad\bm h=\bm a+2\bm Q\bm\mu.
\label{eq:mixture_moments}
\end{equation}
The final trace vanishes for the antisymmetric quadratic form in this two-visit test. Because linear coherent mixing dominates its optical variance, preserving $\bm V$ leaves the leading contribution unchanged. The rare high-variance states can nevertheless dominate the far tail. This construction therefore tests information absent from a variance budget, rather than simply adding more RMS disturbance.

Conditional Poisson counts are retained and the characteristic function of the resulting Gaussian-state-mixture count model is inverted deterministically (Appendix~\ref{app:tails}). Let $q_0$ and $q_1$ be the quantiles of the centered null estimate and planet-present error, respectively. A marginal allocation $\alpha_1=10^{-3}/30000$ controls the search by a union bound without requiring independent hypotheses. The probability-calibrated requirement is
\begin{equation}
f_{\rm th}=q_0(1-\alpha_1),\qquad f_p\ge q_0(1-\alpha_1)-q_1(0.01).
\label{eq:quantile_requirement}
\end{equation}
Equation~\eqref{eq:quantile_requirement} asks whether a threshold exists that rejects sufficient null realizations while retaining 99\% of planet realizations. Increasing the threshold can always reduce false alarms, but may make the planet undetectable. The known optical-bias correction is held fixed in all cases, and distinct null and planet photon variances are retained; uncertain bias remains a separate requirement.

\begin{table}[t]
\caption{Distributional stress test at 80-fm centered single-visit RMS, fixed counts, and the same observing cost. FRN, $P_{\rm FA,1}$, and $f_{\rm th}$ describe the unrecognized-state measurement; $P_{\rm FA,1}$ uses the Gaussian-design threshold 80.016 ppt. The last two columns give the largest ensemble RMS with 99\% average power: $\sigma_{\rm unlab}$ uses a common threshold [Eq.~\eqref{eq:quantile_requirement}], whereas $\sigma_{\rm label}$ assumes the exact state label and the same conditional false-alarm allocation in every state [Eq.~\eqref{eq:conditioned_acceptance}]. No classification uncertainty or additional observing cost is charged to the ideal-label comparator.}
\label{tab:tail_stress}
\begin{tabular}{rrrrrrr}\toprule
$p_e$&$f_v$&FRN (ppt)&$P_{\rm FA,1}$&$f_{\rm th}$ (ppt)&$\sigma_{\rm unlab}$ (fm)&$\sigma_{\rm label}$ (fm)\\\midrule
0&0&14.81825&$3.04\times10^{-8}$&79.77&81.43&81.43\\
.01&.10&14.81830&$1.59\times10^{-4}$&168.01&39.03&67.66\\
.01&.20&14.81847&$6.14\times10^{-4}$&234.10&28.24&62.94\\
.001&.10&14.81885&$2.42\times10^{-4}$&460.64&14.95&85.08\\\bottomrule
\end{tabular}
\end{table}

\begin{figure}[t]
\centering\includegraphics[width=.52\linewidth]{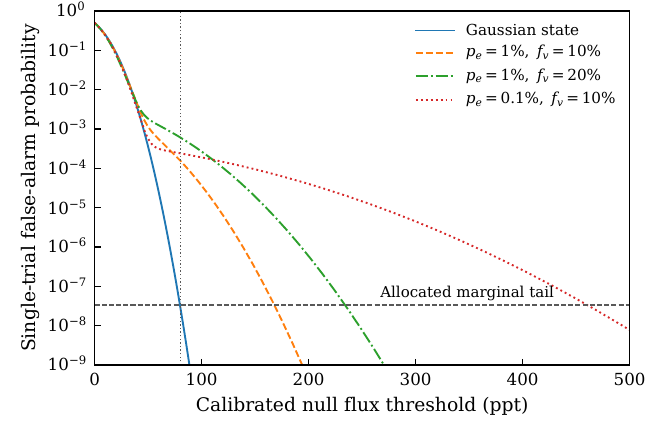}
\caption{Unrecognized-state null survival probabilities for identical disturbance mean, covariance, and 80-fm centered single-visit RMS. The cases of Table~\ref{tab:tail_stress} retain quadratic intensity, conditional Poisson counts, and calibration noise. The horizontal reference is $10^{-3}/30000$; the vertical reference is the 80.016-ppt Gaussian-design threshold. ``Gaussian state'' denotes the disturbance law, not an assumed Gaussian count estimator. Similar FRN conceals very different tail constraints; event frequencies and variance fractions are declared stress inputs.}
\label{fig:tail_stress}
\end{figure}

A one-percent component carrying twenty percent of the variance changes FRN by only 0.0015\%, but raises the false-alarm probability at the Gaussian-design threshold by about $1.8\times10^4$ relative to the allocation (Table~\ref{tab:tail_stress}; Fig.~\ref{fig:tail_stress}). Recalibrating the threshold at the original RMS leaves only $1.11\times10^{-4}$ power. Satisfying both probability objectives instead tightens RMS from 81.43 to 28.24 fm, a factor of 2.88. More accurate covariance estimation alone cannot repair this failure of the assumed tail law.

The contrast between the central distribution and its extreme tail is equally important for validation: ordinary interval coverage need not diagnose the failure. The 1\%/20\% mixture has 95.726\% model coverage inside the nominal Gaussian 95\% interval. An independent $2^{20}$-program planet-present simulation gives 95.720\%, with Wilson interval [95.681,95.759]\%. A separate null simulation gives 637 exceedances in $2^{20}$ programs, consistent with $6.14\times10^{-4}$. These checks concern accessible probabilities; the extreme tail comes from deterministic inversion. The event distribution and prevalence must still be calibrated for an actual observatory.

\paragraph*{State-resolved thresholds and observability.}
The penalty above belongs to a protocol that does not identify the observing state. The same $W$ multiplies all disturbance coordinates, so useful state information may occur in planet-free science samples as well as in telemetry. To separate its potential value from its measurement cost, consider an \emph{ideal-label comparator}: $W=w$ is known exactly before the detection decision, the flux estimator and counts are unchanged, and each state receives the same conditional false-alarm allocation $\alpha_1$. If $F_{h,w}$ is the centered-error CDF under hypothesis $h$, then
\begin{equation}
 f_{{\rm th},w}=F_{0,w}^{-1}(1-\alpha_1),\qquad
 \overline P_D=\sum_w\pi_w\big[1-F_{1,w}(f_{{\rm th},w}-f_p)\big]\ge0.99 .
 \label{eq:conditioned_acceptance}
\end{equation}
The conditional count laws follow from Appendix~\ref{app:tails} with covariance $w\bm V$. Table~\ref{tab:tail_stress} shows the resulting allowances. For the 1\%/20\% mixture, exact labels raise 28.24 to 62.94 fm, reducing the penalty relative to 81.43 fm from 2.88 to 1.29. For the 0.1\%/10\% mixture, 85.08 fm exceeds the Gaussian allowance: the rare state fits largely within the one-percent average miss budget, while quiet programs have lower variance. Neither case requires 99\% power \emph{within} the high state. These are specified conditional tests, not globally optimized allocations of false alarms among states or fundamental tolerance bounds.

Exact labels cannot be inferred merely by adding pixels. The model has only two latent coordinate draws $(x_A,x_B)$ per program; additional samples improve their estimation but do not create independent draws from the state-dependent distribution. The two Gaussian components overlap even with noiseless coordinate estimates. A practical pipeline must propagate the joint state-estimation and flux likelihood, or calibrate its complete selection rule, rather than substitute estimated labels into ideal thresholds. Science data and telemetry are both possible inputs. This distinction complements distribution-aware interferometric detection \cite{Dannert2025NonGaussian} without assuming its instrumental noise law.

\subsection{Physical interpretation and implications for design}
\label{sec:allocation_interpretation}

\paragraph*{Photon supply, suppression, and throughput.}
In the visible benchmark, $C_{\rm leak}/C_p\simeq26.3$: a raw intensity near the planet's flux ratio does not imply comparable aperture count rates, because stellar NI and planet flux use different spatial normalizations. The stellar rate contains $\bar g\Craw$, whereas the planet rate contains $\tau_{\rm core}f_p$. Differencing adds the photon variance of the second background realization, explaining the 95.84\% leakage share of $V_{\rm ph}$. Removing the mean image afterwards cannot remove those arrivals. Optical suppression before detection can reduce this cost, but its leverage weakens when $2C_{\rm leak}\lesssim C_p+2C_b$. At the stated fixed response this crossover is near $1.30\times10^{-11}$ NI, obtained directly from Eq.~\eqref{eq:closed_rates}; it is a photon-budget diagnostic, not a new coronagraph requirement. Zodiacal and exozodiacal surface brightness, aperture size, and detector rates shift it. At the planet-shot-noise limit, further rejection of unrelated light cannot improve the available planet counts.

The throughput derivatives in Sec.~\ref{sec:closure} refer to a change in planet coupling with leakage and background response otherwise held fixed. A common transmission change instead attenuates the planet, stellar leakage, and optical sky together. At fixed flux-ratio floor, in an optical-photon-limited measurement, signal and photon variance both scale with that transmission, so required time scales inversely with transmission, rather than its square. An unchanged detector background restores the inverse-square dependence when it dominates. More generally, the relevant figure is $V_{\rm ph}/[\eta_{\rm live}(C_\star\tau_{\rm core})^2]$, not throughput or raw contrast separately. A new mask, stop, or processing operator must therefore earn its stellar rejection without losing too much planet signal. The channel comparison in Table~\ref{tab:channel_alloc} illustrates the same competition: narrower bins admit fewer photons, while a larger long-wavelength diffraction core admits more diffuse background under the adopted sky law.

\paragraph*{Interference and the spatial scale of an allocation.}
In the coherent-linear regime, Eqs.~\eqref{eq:phase_response} and \eqref{eq:sigma_alloc} give $\sigma_{k,\rm allow}\propto C_{k,\rm alloc}[\Ccoh S_k\mathcal O_k\mathcal P_k(1-\rho_k)]^{-1/2}$ at fixed uncertainty factor. Reducing coherent bias thus lowers both leakage shot noise and sensitivity to subsequent motion, provided the Jacobian and throughput are preserved. Reducing an incoherent halo lowers shot noise but does not remove the retained channel's coherent mixing. Neither benefit is described by a universal WFE-to-contrast conversion. At a linear-response null, the quadratic intensity remains: the scalar model instead gives $\sigma_{\rm allow}\propto(C_{\rm alloc}/S)^{1/2}$ at fixed visit correlation. A favorable phase therefore changes the governing power of disturbance amplitude, not just a numerical prefactor. In an uncontrolled search its maintenance cannot be assumed. Likewise, concentrating a bias and perturbation in the same pixels increases their product and tightens the regional tolerance while raising local leakage. The requirement is set by the fields seen by the estimator, rather than by their separate regional energies.

\paragraph*{Geometric access and radiometric reach.}
For a fixed planet and fixed response, inverse-square dilution reduces planet and stellar leakage rates together. In the leakage-limited regime, signal squared decreases faster than photon variance, giving $T\propto d^2$; a fixed sky/detector rate instead gives $T\propto d^4$. These are the origins of the different distance scalings in Fig.~\ref{fig:target_dependence}. In the constant-color, equal-irradiation family, increasing luminosity moves the orbit outward as $\sqrt{\mathcal L}$ and reduces $f_p$ as $\mathcal L^{-1}$. The planet rate at fixed distance stays constant while stellar leakage rises with $\mathcal L$: angular accessibility improves at the expense of radiometric performance. The adopted fixed calibration error in flux-ratio units adds a further ceiling as the planet contrast falls.

Aperture changes also separate the two limits. For a self-similar pupil and extraction in $\lambda/D$ units, with unchanged normalized response, $C_p$ and $C_{\rm leak}$ scale as $D^2$, while $\bar g$ is unchanged. Uniform diffuse-sky counts scale as $A_{\rm col}\Omega_{\rm core}$ and are approximately independent of $D$ at fixed wavelength. Away from persistent floors, exposure therefore scales as $D^{-2}$ in the stellar/planet-photon regime and as $D^{-4}$ in the uniform-sky regime. At the same time, diffraction gives angular access proportional to $D/\lambda$. These conditional scalings cannot be applied across architectures without updating the pupil, off-axis response, and sky extraction. Longer wavelength reduces phase per unit optical path but enlarges the diffraction pattern; it is not automatically an easier stability or spectroscopy channel.

\paragraph*{What integration and control can repair.}
Writing $\epsilon=s/\FRN_{\rm req}$ in Eq.~\eqref{eq:wall_closure} gives $T=T_{\rm ph}/(1-\epsilon^2)$, where $T_{\rm ph}=V_{\rm ph}/[\eta_{\rm live}(C_\star\tau_{\rm core})^2\FRN_{\rm req}^2]$ is the exposure with zero persistent residual. This form explains the factor 2.64 between the 10- and 13-ppt optical cases: the number of independent photons needed rises sharply as the nonaveraging variance consumes the permitted total. The ceiling is a property of the assumed persistent optical/calibration residual, not of Poisson statistics. By contrast, finite-correlation fluctuations can average down. The visit windows suppress both sufficiently fast fluctuations and sufficiently slow common modes, leaving intermediate temporal power most damaging. For a centered Gaussian OU coordinate the quadratic covariance contains the square of its correlation function, producing the shorter time $\tau_c/2$; its mean fluctuation power still generates photons. Control must consequently be assessed against the observation-weighted spectrum. Greater bandwidth can reject drift while injecting more sensing noise, and extra probing or calibration reduces $\eta_{\rm live}$. Those costs must be charged before a smaller residual is interpreted as a shorter observation.

\paragraph*{Rare states and the scope of verification.}
Linear interference transmits a rare state's larger disturbance into the flux residual, even when reduced quiet-state variance preserves ensemble RMS. The far tail is then controlled by uncommon programs, while ordinary programs control central coverage. Table~\ref{tab:tail_stress} shows that state recognition can recover much of the apparent stability penalty; the recovery depends on the information actually measured, its errors, and any lost observing time. Photon counting and diffraction constrain the chosen measurement, while leakage, repeatability, control cadence, and calibration are partly design choices. The budget identifies which choice changes the limiting mechanism. Table~\ref{tab:products_schema} specifies the architecture-dependent inputs required to apply that decision.

\section{Conclusions}\label{sec:concl}
A coronagraph tolerance is a property of a specified measurement: the admitted photon rates, coherent field response, observing windows, and information used to interpret the data all matter. Building on Roman and HWO error-budget heritage, this work supplies a consistent test of whether a category allocation protects that measurement. It preserves the distinction between photon noise, differential bias, residual covariance, and detection tails rather than replacing them by one contrast number.

The count-derived example shows why the detection objective must precede optical flowdown. At the declared planet flux and search threshold, 20-ppt FRN gives 64.55\% conditional power, whereas 99\% power requires 14.94 ppt. Photon and calibration noise exhaust the optical remainder at 8.11 pc for 100 h under the adopted response: beyond that boundary, tighter stability alone cannot restore feasibility. In the leakage-dominated regime, suppression and planet coupling reduce exposure cost; once other photon terms dominate, further stellar rejection yields diminishing returns. A persistent residual imposes a different limitation because its count uncertainty grows linearly with time and cannot be removed by integration.

Within a feasible photon budget, the modal requirement depends on the projected bias--Jacobian overlap and phase. A response null changes the leading power of disturbance amplitude; finite observing windows change which temporal components survive subtraction. These are physical changes to the measurement, not scalar margins. Control and processing improvements must therefore be evaluated with their sensing noise, planet attenuation, and observing cost included.

The intermittency test establishes that the information retained about the observing state is also part of a requirement. The declared 1\%/20\% mixture leaves FRN nearly unchanged but produces a 2.88-fold tolerance penalty when its state is unrecognized. Exact state labels reduce the penalty to 1.29 for the specified conditional test and average-power objective. Neither factor is a universal hardware margin, and finite state-estimation errors require a joint detection analysis. Architecture-specific optical maps, control histories, and measured calibration and state-distribution laws remain necessary for flight allocation. What is established here is how to test adequacy and identify the limiting remedy, not a numerical HWO flight tolerance.

\section*{Acknowledgments}
The work described here was carried out at the Jet Propulsion Laboratory, California Institute of Technology, Pasadena, California, under a contract with the National Aeronautics and Space Administration.
\textcopyright\ 2026. California Institute of Technology. Government sponsorship acknowledged.

\appendix
\section{Gaussian moments and joint observing states}
\label{app:moments}

Let $\bm\xi=\bm x-\bm\mu$ and $I_q=c_q+\bm a_q^T\bm x+\bm x^T\bm Q_q\bm x$, with real symmetric $\bm Q_q$. Subtracting the mean gives $I_q-m_q=\bm h_q^T\bm\xi+\bm\xi^T\bm Q_q\bm\xi-\Tr(\bm Q_q\bm V)$. Centered Gaussian third moments vanish and
\begin{equation}
\Eavg(\xi_i\xi_j\xi_k\xi_\ell)=V_{ij}V_{k\ell}+V_{ik}V_{j\ell}+V_{i\ell}V_{jk},
\label{eq:gaussian_fourth_moment}
\end{equation}
which proves Eq.~\eqref{eq:gaussian_moments}. For non-Gaussian coordinates, third and fourth centered cumulants add $h_{q,i}Q_{r,jk}\kappa_{ijk}+h_{r,i}Q_{q,jk}\kappa_{ijk}+Q_{q,ij}Q_{r,k\ell}\kappa_{ijk\ell}$ (summation implied). The equality of optical means and covariances does not imply Gaussian intensity or extreme detection tails.

For two possibly different optical states, set
\begin{equation}
\bm z=\begin{pmatrix}\bm x_A\\\bm x_B\end{pmatrix},\quad
\bm V_z=\begin{pmatrix}\bm V_A&\bm V_{AB}\\\bm V_{AB}^T&\bm V_B\end{pmatrix},\quad
\bm a_{\Delta,q}=\begin{pmatrix}\bm a_{A,q}\\-\bm a_{B,q}\end{pmatrix},\quad
\bm Q_{\Delta,q}=\diag(\bm Q_{A,q},-\bm Q_{B,q}).
\label{eq:joint_visit_state}
\end{equation}
Use the stacked mean and $c_{\Delta,q}=c_{A,q}-c_{B,q}$ in Eq.~\eqref{eq:gaussian_moments}. Reference brightness/color rescaling and spectral integration belong in the count operator, not an unjustified subtraction of different NI conventions. Replacing $\bm V$ by the two-time covariance $\bm K(t,t')$ in the same moment identity yields Eq.~\eqref{eq:temporal_quadratic}.

For the integrated test, the signed optical residual before multiplication by $\kappac$ is $y_I=(\Delta I_+-\Delta I_-)/2=\ell(x_A-x_B)+q(x_A^2-x_B^2)$, where $\ell=\Ree[E_0^*(J_+-J_-)]$ and $q=(|J_+|^2-|J_-|^2)/2=0.6S$. Hence $b_{\rm opt}=\kappac\Eavg y_I$ and the optical variance is $\kappac^2\Var(y_I)$. For the four independent conditional Poisson counts, the photon variance is $\sum_{v,q}\Eavg\Lambda_{v,q}/(2t_vC_\star\tau_{\rm core})^2$. 

\section{Count-level implementation verification}
\label{app:countcheck}
For a count-level implementation check, define two nonoverlapping ADI apertures $q=+,-$. Within each visit let the scalar WFE state be constant, so that the experiment explicitly tests the persistent-state limit rather than approximating intra-visit averaging. The exactly linear local fields and an additional static incoherent intensity are
\begin{equation}
E_q(x)=\sqrt{C_0}+J_qx,\qquad
(J_+,J_-)=\sqrt{1.6S}\,e^{i\pi/4}(1,-1/2),\quad
I_q=|E_q|^2+10^{-10},
\label{eq:validation_fields}
\end{equation}
with $C_0=2\times10^{-10}$ NI and $S=6.90\times10^{-7}$ NI nm$^{-2}$. Thus the regional mean Jacobian magnitude is $S$, while the two apertures have different responses to the same mode. The two-visit state is Gaussian with
\begin{equation}
\bm\mu=(0.025,-0.015)\ \mathrm{pm},\qquad
\bm V=(0.080\ \mathrm{pm})^2\begin{pmatrix}1&0.6\\0.6&1\end{pmatrix}.
\label{eq:validation_states}
\end{equation}
Convert to nm in Eq.~\eqref{eq:validation_fields}. Each field denotes a uniform aperture-averaged local response across the visible band; this is an explicitly achromatic budget model, not a fitted HWO optical map. The planet lies in $A,+$ and $B,-$ and is absent in the other two samples. Conditional on $(x_A,x_B)$, draw four independent Poisson counts with means
\begin{equation}
\Lambda_{v,q}=t_v\{C_\star\bar g I_q(x_v)+C_b+C_p\,\delta_{(v,q)\in\{(A,+),(B,-)\}}\},
\quad t_v=\eta_{\rm live}T_{\rm wall}/2.
\label{eq:validation_counts}
\end{equation}
Here the indicator is one at the two planet locations and zero otherwise. All rates, live time, and photon conventions are those of Sec.~\ref{sec:closed_case}; $C_b=0.021$ e$^-$ s$^{-1}$. The calibrated estimator is
\begin{equation}
\widehat f_p=\frac{N_{A,+}-N_{B,+}-N_{A,-}+N_{B,-}}
{2t_v C_\star\tau_{\rm core}}-b_{\rm opt}+\epsilon_{\rm cal},\qquad
\epsilon_{\rm cal}\sim\mathcal N(0,(3.5\ \mathrm{ppt})^2).
\label{eq:validation_estimator}
\end{equation}
The analytically predicted $b_{\rm opt}=6.3729$ ppt is treated as exactly known and subtracted once; this tests model normalization, not achieved calibration accuracy. Before subtraction the RMSE is 16.13 ppt although the FRN is 14.82 ppt. The independent 3.5-ppt calibration residual is one draw per complete observing program, not a term that averages down with frames.

The quadratic moments and the law of total covariance predict photon, optical, and calibration standard deviations of 8.7958, 11.4002, and 3.5000 ppt, respectively. Table~\ref{tab:validation_results} compares the resulting 14.8182-ppt FRN with $2^{20}=1048576$ complete field/count/calibration realizations.  The expected counts, $1.27$--$1.32\times10^5$ per sample, place the photon layer in its nearly Gaussian regime. Gaussian quadrature independently reproduces the analytic optical mean and variance to relative errors below $5\times10^{-13}$.

\paragraph*{Remaining calibration-uncertainty allowance.}
With all modeled terms fixed, an additional zero-mean flux error $e_{\rm add}$ independent of those terms has at most
\[
\sigma_{{\rm add},\max}
=\sqrt{\FRN_{\rm req}^2-\FRN_{\rm model}^2}
=\sqrt{(14.9444\ \mathrm{ppt})^2-(14.8182\ \mathrm{ppt})^2}
\simeq1.94\ \mathrm{ppt}
\]
of standard-deviation allowance. This includes any \emph{additional} random uncertainty in estimating the optical-bias correction, beyond the retained 3.5-ppt calibration term; the two must not be counted twice. A residual correlated with the modeled error adds its cross-covariance term, so this independent allowance does not apply unchanged. A fixed calibration offset instead enters estimator bias and shifts detection probabilities [Eq.~\eqref{eq:estimator_mismatch}]; the 1.94-ppt variance allowance does not certify tolerance to an unknown signed bias.

\begin{table}[t]
\caption{Count-level implementation verification of Eqs.~\eqref{eq:validation_fields}--\eqref{eq:validation_estimator} with 1,048,576 independent complete programs. The full model and the spatial-diagonal surrogate use the \emph{same} simulated data, template, observing time, and marginal aperture variances. Intervals are 95\% Wilson binomial intervals for empirical coverage; Monte Carlo uncertainty is not instrument-parameter uncertainty.}
\label{tab:validation_results}
\begin{tabular}{p{0.39\linewidth}p{0.24\linewidth}p{0.25\linewidth}}
\toprule
Quantity & Full covariance & Spatial-diagonal surrogate \\
\midrule
Predicted FRN (ppt) & 14.8182 & 12.7215 \\
Measured sampling SD (ppt) & 14.8075 & Same estimator/data \\
Mean calibrated error (ppt) & $-0.0020\pm0.0145$ (SE) & Same mean correction \\
Coverage of nominal 95\% interval & 95.015\% & 90.753\% \\
95\% interval on coverage & [94.973, 95.056]\% & [90.698, 90.809]\% \\
\bottomrule
\end{tabular}
\end{table}

The two differential aperture residuals have correlation $-0.9999166$ by construction. The signed extraction therefore amplifies their covariance. In the Gaussian limit the diagonal surrogate's coverage follows directly from
\[
2\Phi_N\!\left(1.959964\,\frac{12.7215}{14.8182}\right)-1=0.907554.
\]
The direction is not universal: changing only $J_-\rightarrow+J_+/2$ gives optical standard deviations 3.8024 ppt (full) and 8.4992 ppt (diagonal). The diagonal model is then 2.235 times conservative for the optical term. Including photon and calibration terms gives 10.2023 versus 12.7226 ppt, a ratio of 1.247 for total FRN. This controlled sign reversal tests covariance bookkeeping, not the distributional or optical assumptions. Those are tested separately in Sec.~\ref{sec:validation}.

\section{Numerical inputs and convergence}
\label{app:numerical}

Use $c=299792458$ m s$^{-1}$, $h=6.62607015\times10^{-34}$ J s, $k_B=1.380649\times10^{-23}$ J K$^{-1}$, $1$ au$=149597870700$ m, and $1$ pc$=(648000/\pi)$ au. Integrate Eqs.~\eqref{eq:blackbody_rate}--\eqref{eq:band_geometry} over $\lambda_c(1\pm\Delta\lambda/(2\lambda_c))$ with common efficiency 0.20 and the angular aperture in Table~\ref{tab:closure_inputs}. Set $C_{\rm leak}=C_\star\bar g\Craw$, $C_p=C_\star\tau_{\rm core}f_p$, and $C_b$ to sky plus dark. The positive optical remainder is $\sqrt{\FRN_{\rm req}^2-\FRN_{\rm rand}^2-\FRN_{\rm cal}^2}/\kappac$; the channel table instead reports photon-plus-calibration precision and continuum SNR. Minimum times set the optical residual to zero in Eq.~\eqref{eq:wall_closure}.

The frequency-domain OU check integrates in $u=fT$, resolving the oscillatory sinc/window factors rather than treating the tail as smooth. On $0\le u\le32$ use adaptive quadrature with absolute tolerance $10^{-12}$ and relative tolerance $10^{-10}$; from 32 to 4096 use 24-node Gauss--Legendre quadrature on half-unit intervals. For adjacent visits the omitted dimensionless variance tail is bounded by $4/(3\pi^4rF^3)$ at cutoff $F=4096$. For large $r$, evaluate $A(r)=1-(3r)^{-1}+(12r^2)^{-1}-(60r^3)^{-1}+\cdots$ to avoid cancellation. The analytic and numerical checks in Sec.~\ref{sec:completed_checks} refer to this specified process, not to measured control spectra.

Monte Carlo samples use NumPy's PCG64 pseudorandom-number generator, initialized once with seed 20260921, in double precision and 64 batches of 16384 programs. In each batch draw row-major standard normals $Z$ of shape \texttt{(16384,2)} and form $\bm\mu+ZL^T$, where $L$ is the lower Cholesky factor of Eq.~\eqref{eq:validation_states}. Evaluate intensities with axes (program, visit $A/B$, aperture $+/-$). Draw the complete Poisson array in one call, equivalently \texttt{(16384,4)} ordered $A+,A-,B+,B-$; then draw 16384 calibration errors and apply Eq.~\eqref{eq:validation_estimator}. Regrouping draws changes only the Monte Carlo digits. Batch-variance dispersion divided by eight is 0.315 ppt$^2$; 12- and 16-node Gaussian quadrature checks the optical moments.

For the intermittent checks use seeds 20260922 (null) and 20260923 (planet), with the same batching. Before each normal array, draw 16384 independent $u\sim\mathcal U[0,1)$ variates; assign the high state $W=f_v/p_e$ exactly when $u<p_e$, and the quiet state $W=(1-f_v)/(1-p_e)$ otherwise. Multiply Gaussian deviations by $\sqrt W$ before the Poisson and calibration draws. Each hypothesis has $2^{20}$ programs. Simulations check central coverage and the accessible $\sim10^{-4}$ tail; Appendix~\ref{app:tails} supplies extreme-tail probabilities.

Coverage uses $|\widehat f_p-f_p|\le1.959964\sigma_{\rm model}$ with fixed model uncertainties and the covariance comparison of Appendix~\ref{app:countcheck}. For $k$ covered realizations out of $n$, Wilson intervals use
\begin{equation}
\frac{\widehat p+z^2/(2n)\ \pm\ z\sqrt{\widehat p(1-\widehat p)/n+z^2/(4n^2)}}{1+z^2/n},\qquad \widehat p=k/n,\quad z=1.959964.
\label{eq:wilson_interval}
\end{equation}
The intervals quantify Monte Carlo precision, not instrumental uncertainty or the accuracy of extreme detection tails.

\section{Exact count-distribution inversion}
\label{app:tails}
Let $D=N_{A,+}+N_{B,-}-N_{A,-}-N_{B,+}$, with centered error in ppt $e=r_ND-c_e+\epsilon_{\rm cal}$, $r_N=10^{12}/(2t_v\nu)$, and $\nu=C_\star\tau_{\rm core}$. Here $c_e=b_{\rm opt,ppt}+f_{p,\rm ppt}$ under the planet hypothesis; omit the planet term under the null. The two independent conditional Poisson sums have quadratic means
\begin{equation}
\Lambda_\pm(\bm x)=c_\pm+\bm b_\pm^T\bm x+\bm x^T\bm Q_\pm\bm x.
\label{eq:poisson_quadratic_means}
\end{equation}
For $a_q=2\Re(E_{0,q}^*J_q)$, $q_q=|J_q|^2$, and $C_I=C_\star\bar g$,
\begin{align}
c_+&=2t_v[C_I(C_0+10^{-10})+C_b]+2t_vC_p,&c_-&=2t_v[C_I(C_0+10^{-10})+C_b],\nonumber\\
\bm b_+&=t_vC_I(a_+,a_-)^T,&\bm b_-&=t_vC_I(a_-,a_+)^T,\nonumber\\
\bm Q_+&=t_vC_I\diag(q_+,q_-),&\bm Q_-&=t_vC_I\diag(q_-,q_+).
\label{eq:poisson_coefficients}
\end{align}
Set $C_p=0$ under the null. With $v_\pm(t)=e^{\pm it r_N}-1$, form $c_t=v_+c_++v_-c_-$, $\bm b_t=v_+\bm b_++v_-\bm b_-$, and $\bm Q_t=v_+\bm Q_++v_-\bm Q_-$. For a component with covariance $w\bm V=\bm L_w\bm L_w^T$, define
\begin{equation}
\bm A_w=\bm I-2\bm L_w^T\bm Q_t\bm L_w,\qquad\bm d_w=\bm L_w^T(\bm b_t+2\bm Q_t\bm\mu).
\label{eq:cf_matrices}
\end{equation}
Completing the Gaussian square gives
\begin{equation}
\phi_e(t)=e^{-itc_e-\frac12(3.5t)^2}\sum_w\pi_w
\frac{\exp[c_t+\bm b_t^T\bm\mu+\bm\mu^T\bm Q_t\bm\mu+\frac12\bm d_w^T\bm A_w^{-1}\bm d_w]}
{\sqrt{\det\bm A_w}}.
\label{eq:exact_count_cf}
\end{equation}
Use the root continuous from unity at $t=0$, ordinary transposes, and mixture weights from Eq.~\eqref{eq:mixture_state} ($w=1$ for Gaussian states). Gil--Pelaez inversion \cite{GilPelaez1951Inversion} gives
\begin{equation}
P(e>x)=\frac12+\frac1\pi\int_0^\infty\frac{\Im[e^{-itx}\phi_e(t)]}{t}\,dt.
\label{eq:cf_inversion}
\end{equation}
The calculation integrates $0\le t\le3$ ppt$^{-1}$ in intervals of width 0.025 ppt$^{-1}$, with 24-node Gauss--Legendre quadrature per interval. Independent Gaussian calibration supplies a damping factor, bounding the omitted tail by $e^{-\sigma_{\rm cal}^2t_{\max}^2/2}/(\pi\sigma_{\rm cal}^2t_{\max}^2)<4\times10^{-27}$. This is separate from quadrature and floating-point error. Refinement to 12 or 32 nodes changes the 1\%/20\% threshold by less than $10^{-5}$ ppt; cutoffs 2.5 and 3.5 ppt$^{-1}$ give the same printed value. 

For Eq.~\eqref{eq:nonlinear_completion}, use the same joint Gaussian states, 16/32 quadrature nodes per coordinate, and RMS scales $(0.008105,0.01351,0.04870,0.080,0.20,1.0)$ pm, converted to nm.

\paragraph*{Known-state calculation.}
For Eq.~\eqref{eq:conditioned_acceptance}, evaluate Eq.~\eqref{eq:exact_count_cf} one component at a time with covariance $w\bm V$, keeping the mean, photon response, and bias correction fixed. Root-solve each conditional null quantile and then the ensemble RMS giving $\sum_w\pi_w P_{D|w}=0.99$. Sixteen- and thirty-two-node inversion agrees within $10^{-4}$ fm. At the 1\%/20\% allowance, quiet- and high-state powers are 0.999954 and 0.004563: high-state completeness is deliberately not constrained to 99\%. The comparison uses exact labels at no extra cost and does not optimize false-alarm allocations across states.


%

\end{document}